\documentclass[aps,prb,twocolumn,superscriptaddress,longbibliography]{revtex4-2}

\usepackage{array}
\usepackage{pgfplots}
\usepackage{amsfonts}
\usepackage[colorlinks=true]{hyperref}
\usepackage{amssymb}
\usepackage{mathrsfs}
\usepackage{amsmath}
\usepackage{amsthm}
\usepackage{bm}
\usepackage{amsfonts}
\usepackage{dsfont}
\usepackage{graphicx}
\usepackage{lipsum}
\usepackage{mathtools}
\usepackage{physics}
\usepackage{algorithm}

\usepackage[normalem]{ulem}
\usepackage{color}
\usepackage{xcolor}

\newcommand{\bd}[1]{\boldsymbol{#1}}

\newcommand{\wb}{\boldsymbol{w}}

\definecolor{Ugreen}{HTML}{198a11}

\begin{document}

\title{Refining the weighted subspace-search variational quantum eigensolver:  compression of ans\"atze into a single pure state and optimization of weights
}

\author{Cheng-Lin Hong}
\affiliation{Faculty of Physics, Arnold Sommerfeld Centre for Theoretical Physics (ASC),\\Ludwig-Maximilians-Universit{\"a}t M{\"u}nchen, Theresienstr.~37, 80333 M{\"u}nchen, Germany}
\affiliation{Munich Center for Quantum Science and Technology (MCQST)}

\author{Luis Colmenarez}
\affiliation{Institute for Quantum Information, RWTH Aachen University, 52056 Aachen, Germany}
\affiliation{Institute for Theoretical Nanoelectronics (PGI-2), Forschungszentrum Jülich, 52428 Jülich, Germany}
\affiliation{Max Planck Institute for the Physics of Complex Systems, N\"othnitzer Stra{\ss}e~38, 01187-Dresden, Germany}

\author{Lexin Ding}
\affiliation{Faculty of Physics, Arnold Sommerfeld Centre for Theoretical Physics (ASC),\\Ludwig-Maximilians-Universit{\"a}t M{\"u}nchen, Theresienstr.~37, 80333 M{\"u}nchen, Germany}
\affiliation{Munich Center for Quantum Science and Technology (MCQST)}

\author{Carlos L. Benavides-Riveros}
\email{cl.benavidesriveros@unitn.it}
\affiliation{Max Planck Institute for the Physics of Complex Systems, N\"othnitzer Stra{\ss}e~38, 01187-Dresden, Germany}
\affiliation{Pitaevskii BEC Center, CNR-INO and Dipartimento di Fisica, Università di Trento, I-38123 Trento, Italy}

\author{Christian Schilling}
\email{c.schilling@physik.uni-muenchen.de}
\affiliation{Faculty of Physics, Arnold Sommerfeld Centre for Theoretical Physics (ASC),\\Ludwig-Maximilians-Universit{\"a}t M{\"u}nchen, Theresienstr.~37, 80333 M{\"u}nchen, Germany}
\affiliation{Munich Center for Quantum Science and Technology (MCQST)}

\date{\today}

\begin{abstract}
The weighted subspace-search variational quantum eigensolver (SSVQE) is a prominent algorithm for calculating excited-state properties of molecular quantum systems. In this work, we elaborate on some of its fundamental features with the aim of improving its practical realization. First, we demonstrate that the initial ans\"atze for various excited states could be prepared into a single pure state through a minimal number of ancilla qubits, followed by the optimization of a subsequent global unitary rotation in the targeted subspace. Since the ancillas' sole purpose is to purify an underlying ensemble $\rho_{\wb}$ state with spectral weights $\wb$, their measurement would just collapse $\rho_{\wb}$ with probabilities $w_j$ to one of its eigenstates $\ket{\Psi_j}$. We thus observe that our realization of SSVQE is equivalent to the original SSVQE improved by importance sampling. Then, we elaborate by numerical means on the potential influence of the auxiliary weights $\wb$ on the accuracy of the sought-after eigenstates and eigenenergies. Clear trends are discovered which are contrasted with some recent mathematical results concerning the ensemble variational principle that underlies SSVQE.
\end{abstract}

\maketitle

\section{Introduction}

Describing and understanding excited-state properties of atoms, mo\-le\-cu\-les, and solids, is one of the most important challenges of modern electronic structure theory. While state-of-the-art methodologies can reach a great degree of accuracy for molecular ground states at equilibrium, excited states present themselves as a major computational bottleneck for at least two reasons: the many types of qualitatively different electronic excitations present in chemical processes (valence, Rydberg, or vertical, to list a few) and their inherent stronger multi-con\-fi\-gu\-ra\-tio\-nal character, expanding thus into quite large parts of the Hilbert space \cite{Gonzalez2012,RevModPhys.74.601,SERRANOANDRES200599,Westermayr,Loos2018,Ghosh2018}. This also makes their manipulation and storage very inefficient on classical computers \cite{book2020}. A promising strategy to overcome some of these problems is to use quantum devices, whose dis\-rup\-tive potential lies precisely in their capability of representing and manipulating quantum states in such challenging physical setups \cite{Cirac2012,Preskill2018quantumcomputingin}. The possibility of controlling highly multiconfigurational wave functions in near-term quantum technologies has recently made the development of hybrid quantum-classical algorithms for excited states an active field of research \cite{OSRVE,PhysRevLett.122.230401,PhysRevResearch.2.043140,Lyu2023symmetryenhanced,PhysRevX.8.011021,Gocho2023,tazi2023folded,Review2021}.

Hybrid quantum-classical algorithms were first developed for chemical ground states. Chief among them is the variational quantum eigensolver (VQE) \cite{Peruzzo2014}.  It prepares trial wave functions on a quantum circuit and optimizes their parameters self-consistently using classical devices, requiring short coherence times in the execution of the circuit \cite{McClean_2016,Fedorov2022}.
In virtue of generalized variational principles,  several approaches were proposed for calculating excited states in the realm of variational and contracted quantum eigensolvers  \cite{ChemRev,doi:10.1126/sciadv.aap9646,RevModPhys.92.015003,PhysRevA.95.042308,Zhang_2022,wang2023electronic,Benavides-Riveros_2024,Wen2024fullcircuitbased}. Two of the most prominent, yet conceptually different ones, are the variational quantum deflation (VQD) algo\-rithm and the subspace-search VQE (SSVQE). VQD targets excited states  \emph{successively}: the $K$-th excited state is calculated variationally from the original Hamiltonian modified by suitable energy penalties for the first $K$ eigen\-states $\ket{0}, \ldots ,\ket{K-1}$, determined beforehand \cite{Higgott2019variationalquantum,PhysRevA.99.062304,PhysRevResearch.4.013173,Wen2021,Shirai}. In contrast, SSVQE identifies in the first place just the subspace spanned by the $K$ lowest-lying eigenstates. For this, a unitary is sought which rotates  $K$ orthogonal initial states in such a way that the \emph{average} energy of the resulting states is minimized \cite{PhysRevResearch.1.033062,Yalouz_2021,PhysRevResearch.6.013015}.

While the significance of VQD and SSVQE in the context of quantum computing of excited states can hardly be overestimated, they are both plagued by some unpleasant limitations and shortcomings. For instance, to determine the $K$-th eigenstate, VQD needs to calculate each of the first $K\!-\!1$ eigenstates in separate, preceding calculations, and, as $K$ increases, its numerical predictability deteriorates due to an intrinsic accumulation of errors \cite{PhysRevResearch.3.013197}. Those deficiencies are overcome by SSVQE in the sense that the or\-tho\-gonality is fulfilled automatically.  The resulting individual eigenstates, however, are optimal only on average, without control over the errors, and their calculation in the first place required additional classical or quantum resources. This has motivated a modification of SSVQE which assigns positive weights $\boldsymbol{w}=(w_0,..,w_{K-1})$ to the targeted eigenstates in the subspace. Yet, while this makes supplementary classical (or quantum) optimizations within the subspace obsolete, the impact of the choice of those auxiliary weights on the quality of the targeted eigenenergies and eigenstates has not been explored yet.

It will be the ultimate accomplishment of our work to refine some fundamental aspects of weighted SSVQE to improve its practical realization. First, we show that the initial ans\"atze for various excited states can be prepared into a single pure state through a minimal number of ancilla qubits. As a result, various excited states can be determined simultaneously, rather than individually in separate calculations. We then observe that measuring in addition the ancillas would just collapse the variational ensemble state $\rho_{\wb}$ to its eigenstates $\ket{\Psi_j}$ with probabilities $w_j$. Accordingly,  our realization as a purified SSVQE is equivalent to the original SSVQE improved by importance sampling. The latter would actually reduce the number of samples compared to the original deterministic sampling of the trial states.
Second, we study the influence of the auxiliary weights used in SSVQE on the accuracy of the sought-after eigenstates and eigenenergies and discuss strategies for choosing such weights in order to optimize the number of iterations used in the algorithm.

The structure of the paper is as follows. In Sec.~\ref{sec2}, we recap in more detail hybrid quantum-classical methods for excited states and their shortcomings. The latter motivates Sec.~\ref{sec3} where we present our approach on a conceptual level and compare its computational cost to other implementations of SSVQE. In Sec.~\ref{sec4}, we discuss practical aspects and present numerical experiments together with an analysis of the effect of noise on our results. Sec.~\ref{secw} presents our numerical analysis of the role of the weights on the quality of the eigenstates and eigenenergies when using a gradient-based algorithm. Additional technical details are presented in additional appendices.

\section{Hybrid quantum-classical methods for excited states}
\label{sec2}

In this section, we recall in more detail the most prominent quantum eigensolvers established in recent years for computing excited states. A particular emphasis lies on their strengths and weaknesses which in turn motivate our subsequent sections. In the original VQE algorithm, a set of variational parameters $\bd{\theta}$  is optimized on a classical device. These parameters are used to parameterize a quantum circuit and a respective unitary $U(\bd{\theta})$ that transforms an initial reference state $\ket{\phi}$ to a quantum state $\ket{\psi(\bd{\theta})}=U(\bd{\theta}) \ket{\phi}$. To calculate the ground state $\ket{\psi(\bd{\theta}^\ast)}$ of a given Hamiltonian $\hat{H}$, the optimal parameters $\bd{\theta}^*$ are thus found by minimizing the corresponding energy expectation value
$E(\bd{\theta}) =\bra{\psi(\bd{\theta})} \hat{H} \ket{\psi(\bd{\theta})}$.
By exploiting the orthogonality of the eigenstates of $\hat{H}$, VQD adds a sequence of penalty terms to the VQE cost function $E(\bd{\theta})$ to estimate excited states \cite{Higgott2019variationalquantum}.
The effective Hamiltonian for such an optimization of the $K$-th excited state is given by
\begin{align}
\label{eq:VQD}
\hat{H}_K(\bd{\theta}_0,\ldots ,\bd{\theta}_{K-1})  = \hat{H} + \sum^{K-1}_{j=0} \beta_j \ket{\psi_j(\bd{\theta}_j)}\bra{\psi_j(\bd{\theta}_j)}\,.
\end{align}
For sufficiently large positive values $\beta_j$, this is equivalent to minimizing the energy of the $K$-th excited state $E_K(\bd{\theta}_K) =\bra{\psi_K(\bd{\theta}_K)} \hat H \ket{\psi_K(\bd{\theta}_K)}$ subject to the constraint that the minimizer is orthogonal to the states $\ket{\psi_0(\bd{\theta}_0)}$,\ldots, $\ket{\psi_{K-1}(\bd{\theta}_{K-1})}$. The $K$-th eigenstate becomes the ground state of the respective effective Hamiltonian \eqref{eq:VQD}. While there are ways to determine the magnitude of the penalty terms $\beta_0,\ldots,\beta_{K-1}$, leading to faster convergence in the optimization of $\bd{\theta}$
\cite{PhysRevResearch.3.013197}, VQD requires in addition all an\-s\"atze $\ket{\psi_j(\bd{\theta}_j)}$ to be sufficiently expressive. As a result, a progressive accumulation of errors seems to be unavoidable, especially for the calculation of highly excited states. In a rather similar fashion, the orthogonal state reduction VQE (OSRVE) computes the $K$-th state by eliminating contributions of the lower eigenstates \cite{OSRVE}. Instead of requiring penalty terms, OSRVE enforces orthogonality through auxiliary qubits that are added to the circuit. Yet, in complete analogy to VQD, OSRVE requires $K$ rounds of optimization to compute the $K$ energetically lowest eigenenergies and the numerical predictability deteriorates as $K$ increases.

In contrast to VQD and OSRVE, SSVQE in its original form seeks in a first step just the subspace spanned by the first $K$ eigenstates. This is realized in practice through the following \emph{state-a\-ve\-rage} calculation. One chooses $K$ initial orthogonal states $\ket{\phi_0},\ldots,\ket{\phi_{K-1}}$, which are then rotated through a unitary $U(\bd{\theta})$ in order to minimize their energy average. This ``de\-mo\-cratic'' distribution of states results in the corresponding cost function \cite{PhysRevResearch.1.033062,Yalouz_2021,PhysRevA.107.052423}:
$\mathcal{L}(\bd{\theta})=\sum^{K-1}_{j=0} \bra{\phi_j}U^\dagger(\bd{\theta}) \hat{H} U(\bd{\theta})\ket{\phi_j}$.
While the orthogonality of various optimized $U(\bd{\theta}^\ast)\ket{\phi_j}$ is guaranteed, these individual states are optimal only \emph{on average}. Indeed, since all states possess equal weight, this cost function is invariant under unitary rotations among them. Hence, an additional optimization (classical) scheme within this subspace is necessary to extract the individual eigenstates. An advanced variant of SSVQE assigns \emph{distinct} weights $w_j$ to the subspace states in order to make these supplementary optimization stages obsolete.
Then, the generalization of the Rayleigh-Ritz variational principle to ensemble states with spectrum $\bd{w} = (w_0,\ldots,w_{K-1})$, where $w_j > w_{j+1}$ \cite{PhysRevA.37.2809}, guarantees that the energy expectation is bounded from below by the weighted sum of the eigenenergies,
\begin{align}
\sum^{K-1}_{j=0} w_j \bra{\phi_j}U^\dagger(\bd{\theta}) \hat{H} U(\bd{\theta})\ket{\phi_j} \geq \sum^{K-1}_{j=0} w_j E_j\,,
\label{costSSVQE}
\end{align}
with $E_j \leq E_{j+1}$ being the exact eigenenergies of the system.  This ensemble variational principle, which provides the theoretical foundation of weighted SSVQE, offers a unified variational approach to quantum mechanics. The variational principle for ground states is, in fact, just a particular case, corresponding to  $\bd{w} = (1,0,0,\ldots)$. The weights $w_j$ are, in theory, only \textit{auxiliary} parameters, meaning the validity of the ensemble variational principle \eqref{costSSVQE} is independent of the choice of $\bd{w}$. In practice, however, the choice of $\bm{w}$ can have a notable impact given that one never reaches exact mathematical convergence. Specifically, the original work \cite{GOK} on the ensemble variational principle does not elaborate on the crucial relation between the error of the ensemble energy, and those of the physical quantities of interest. This last missing piece of the puzzle has been provided recently by Ref.~\cite{ding2024ground}, where tight bounds on the errors of the sought-after eigenenergies and eigenstates were derived. 

Some of the potential advantages of the weighted SSVQE are the following ones. Relative to VQD and OSRVE, there is no need anymore to enforce the orthogonality constraint of the states through penalty terms. Moreover, the optimization of the quantum circuit runs only once, again in stark contrast to VQD and OSRVE which both require $K$ subsequent optimizations. Finally, the spectrum can be exactly reached as long as the quantum circuit has sufficient expressive power to represent the unitary which maps the input states $\ket{\phi_j}$ to the sought-after eigenstates $\ket{\psi_j(\bd{\theta}^*)}=U(\bd{\theta}^*)\ket{\phi_j}$ of $\hat{H}$. However,  measuring each term $\bra{\phi_j(\bd{\theta})}\hat H\ket{\phi_j(\bd{\theta})}$ with the same number of samples will simply increase the computational cost of SSVQE $K$ times with respect to the ground state algorithm
\cite{Kawai_2020,Zhang_2022,Lyu2023symmetryenhanced}. From a more general conceptual perspective, neither weighted SSVQE nor any of the other discussed VQEs prepare a quantum state that contains information about various excited states \emph{simultaneously}, which could be necessary to determine in a straightforward manner physically relevant quantities, i.e., transition amplitudes and energy gaps. Moreover, as far as the performance of all these algorithms is concerned, the problem of violating symmetries is even more crucial for excited states than for ground states \cite{PhysRevA.101.052340,Lyu2023symmetryenhanced}. Symmetries in VQE-based algorithms are usually enforced so far either through circuits that respect these symmetries or by adding extra penalty terms to the cost function.  While the former strategy could be rather unpractical \cite{Lyu2020accelerated}, the latter requires prior knowledge of various symmetries \cite{Ryabinkin2019,PhysRevResearch.3.013197}.
In summary, while all these algorithms represent a landmark in the development of quantum computation for excited states, they leave room for improvement. 

\section{Quantum computation of excited states by ans\"atze compression}\label{sec3}

Our refined variant of the SSVQE algorithm calculates the weighted cost function
\begin{eqnarray}
\label{QPVQEcost}
\mathcal{L}_{\bd{w}}(\bd{\theta})\equiv \sum^{K-1}_{j=0} w_j E_j(\bd{\theta}) &=&\sum^{K-1}_{j=0} w_j \bra{\phi_j}U^\dagger(\bd{\theta}) \hat{H} U(\bd{\theta})\ket{\phi_j} \nonumber \\
&=& \mbox{Tr}[\hat{H} U(\bd{\theta})\rho(\bm{w})U^\dagger(\bd{\theta})]
\end{eqnarray}
in a direct manner through the mixed quantum state
\begin{align}
\label{mixed}
\rho(\bm{w}) = \sum_{j=0}^{K-1} w_j \ket{\phi_j}\bra{\phi_j}\,,
\end{align}
which satisfies the normalization condition $\sum_j w_j = 1$.
$\mathcal{L}_{\bd{w}}(\bd{\theta})$ is also called \emph{ensemble energy}. In order to achieve this on a unitary quantum circuit, the underlying key idea is to map this state into a single quantum state
\begin{align}
\label{wfield}
\rho(\bm{w}) \xrightarrow{\mathrm{purification}} \ket{\Phi(\bm{w})} = \sum^{K-1}_{j=0} \sqrt{w_j} \,\ket{\phi_j}\otimes\ket{a_j}\,.
\end{align}
The auxiliary states $\ket{a_j}$ are added to execute the parallelization of the eigenstates. The only formal requirement
on $\ket{a_i}$ is the orthonormality condition $\langle {a_i}\ket{a_j} = \delta_{ij}$. As a consequence, $\langle \Phi(\bm{w}) \ket{\Phi(\bm{w})} = 1$, and more generally one finds that for any observable $\hat A$:
$\bra{\Phi(\bm{w})} \hat A \otimes \openone  \ket{\Phi(\bm{w})}  = \Tr[\hat A \rho(\bd{w})]$. Then, $\ket{\Phi(\bm{w})}$ can be used to recover the cost function \eqref{QPVQEcost} which in turn can be rewritten as
\begin{align}
\label{pssveqcost}
\mathcal{L}_{\bd{w}}(\bd{\theta})=\bra{\Phi(\bm{w})}[U^\dagger(\bd{\theta}) \hat H  U(\bd{\theta})]\otimes \openone  \ket{\Phi(\bm{w})}\,.
\end{align}

 Using the state \eqref{wfield} as the starting point for the variational calculation of the parameters through the cost function $\mathcal{L}_{\bd{w}}(\bd{\theta})$ \eqref{pssveqcost} has the following advantages.

 In the first place, an optimal operator averaging is achieved. To explain this, let us recall that for $K$ non-zero weights $\bd{w}$, the standard version of SSVQE refers to an initial computational (orthonormal) basis $\{|\phi_j\rangle\}$. The parameters of the unitary $U(\bd{\theta})$ are optimized such that
 $U(\bd{\theta})$ maps the initial basis states to good approximations of the lowest $K$ eigenstates of a Hamiltonian $\hat H$. In particular, for a given set of parameters $\bd{\theta}$, the individual energies $E_j(\bd{\theta}) = \bra{\phi_j(\bd{\theta})}\hat H\ket{\phi_j(\bd{\theta})}$ shall be determined accurately through $M_j$ many runs/energy measurements for various $j=0,1,\ldots, K-1$. From a general point of view, there are different strategies to choose those $M_j$.
 For example, (a) an equal number of measurements $M_0=\ldots = M_{K-1}$, (b) distinctive $M_j$ such that every energy $E_j(\bd{\theta})$ has the same (relative) error, (c) optimizing the number of samples $M_j$ for each state to achieve a fixed ensemble energy $\mathcal{L}_{\bd{w}}$ error, or (d) sampling the states according to the probability distribution $\{w_0,\ldots,w_{K-1}\}$. The first choice (a) is a suboptimal deterministic sampling, resulting in run-time overhead \cite{Kawai_2020,Zhang_2022,Lyu2023symmetryenhanced}.
 As explained in Appendix \ref{appd}, strategy (c) is more efficient than strategy (a) and (b) in the sense that in order to achieve the same ensemble energy error a smaller total number of shots is needed. To be more specific, strategy (a) increases the number of shots with respect to strategy (c) by a factor $K f(\bd{w},\bd{\theta})\geq 1$, where $f(\bd{w},\bd{\theta}) = \sum_j w^2_j \sigma^2_j(\bd{\theta})/\big[\sum_j w_j \sigma_j(\bd{\theta})\big]^2$ and $\sigma_j^2(\bd{\theta})$ are the individual statistical variances of the energy.
 On the other hand, strategy (d), i.e., SSVQE improved by importance sampling, optimizes the number of shots for random, unbiased estimations of the energy while achieving the same accuracy as the more costly deterministic sampling (c) \cite{https://doi.org/10.1002/wics.56,Kiss2023importancesampling}.


 Interestingly, our approach implicitly realizes the method of \textit{importance sampling}, where the quantum states are sampled according to a classical probability distribution determined by the weights $\bd{w}$. To see this connection, we first observe that measuring the ancilla qubits in \eqref{wfield} would collapse the working qubits to the state $|\phi_j\rangle$ with probability $w_j$. Now, measuring the energy on the working qubit would yield the expectation $\langle \phi_j|\hat{H}|\phi_j\rangle$. Taking into the conditional probability of the outcome of the measurement on the ancilla, we find that the final expectation value of the energy is not affected by this additional measurement step
 \begin{equation}
     \langle \hat{H} \rangle = \sum_i w_i \langle \phi_j |\hat{H} |\phi_j\rangle = \mathrm{Tr}[\rho(\bd{w})\hat{H}].
 \end{equation}
 The same reasoning holds true if one first applies a unitary $U(\bd{\theta})$ on the working qubits since the measurement of the ancillas will not affect them. Based on this, it is becomes clear that the measurement performed on the ancillas resembles nothing else than a random number generator used in importance sampling: The measurement on the ancilla correctly ``prepares'' the working qubits in state $|\phi_j\rangle$ and $U(\bd{\theta})|\phi_j\rangle$, respectively, with probability $w_j$. Therefore our refined approach can be viewed as a purely quantum version of importance sampling, where the probability distribution is supplied by the laws of quantum mechanics.

Second, despite involving the simultaneous rotation of various eigenstates, the cost function can be optimized by using only one quantum circuit, which reduces the time for circuit reset and compiling, in contrast to, e.g., weighted SSVQE which requires setting up and compiling $K$ different circuits, and summing up the individual energies $E_j(\bd{\theta})$. Finally, enforcing a given symmetry $\hat s$ in the cost function is also possible in our scenario: instead of computing a cost function for each individual state as it is usually done both in VQE and VQD,
one can utilize  $\bra{\Phi'(\bm{w})} \mathcal{U}^\dagger (\bd{\theta})\hat{S} \,\mathcal{U}(\bd{\theta})\ket{\Phi'(\bm{w}}$, where  $ \hat S = \hat s^2\otimes \openone \otimes \openone - \hat s \otimes \hat s \otimes \openone $, $\mathcal{U} (\bd{\theta}) = U(\bd{\theta})\otimes U(\bd{\theta})\otimes \openone$,  and $\ket{\Phi'(\bm{w})} = \sum_j \sqrt{w_j} \,\ket{\phi_j}\otimes\ket{\phi_j} \otimes\ket{a_j}$. The latter is a generalization of the state \eqref{wfield}. This way of computing fluctuation of observables (i.e., by doubling the Hilbert space) has already prominently been used in the context of tensor network ans\"atze for quantum spectra \cite{PhysRevB.94.041116}. For quantum devices, the preparation of this kind of state has been discussed in Refs.~\cite{Sagastizabal2021,PhysRevResearch.4.013003,PhysRevLett.123.220502}.

Because of these promising features, we refer to our proposed approach in the following as \textit{purified} SSVQE.

\section{Implementation and numerical experiments}
\label{sec4}

\subsection{Initial state preparation}
\label{sec:fixing_particle_number}
To describe the state preparation of our purified SSVQE, we start with  $M$ real value orthogonal spatial orbitals $\ket{f_p}$, with $p=0,\ldots,M-1$. Under the spin-restricted formalism, we define the corresponding $\alpha$ ($\beta$) spin-orbitals denoted as $\ket{f_p} \ket{\alpha} $ (and $\ket{f_p} \ket{\beta} $) for each spatial orbital, resulting in a total of $N=2M$ spin orbitals. In the Jordan-Wigner (JW) encoding \cite{Jw}, the occupation number of a spin-orbital is represented by the states $\ket{0}$ or $\ket{1}$ of a corresponding qubit state (unoccupied and occupied, respectively). An `interleaved' ordering is utilized to assign numbers to the qubits such that  $\ket{f_{0\alpha}},\ket{f_{0\beta}}, \ldots, \ket{f_{(M-1)\alpha}},\ket{f_{(M-1)\beta}}$.
The reference states $\ket{\phi}$ used in Eq.~\ref{wfield} have fixed particle numbers, with $n_{\alpha}$ (the number of spin-up electrons) equal to $n_{\beta}$ (the number of spin-down electrons). These reference states represent Slater determinants with a total spin projection of $S_z=0$.

The aim is to prepare an initial quantum state reading $\sum_{j=0}^{K-1} \sqrt{w_j} \ket{D_j}$, where $\ket{D_j}$ denotes a computational basis state, corresponding to a bitstring that encodes the occupation of the $N$ spin-orbitals under JW encoding.  Different strategies can be employed to achieve this task \cite{Qubitstatevector,QSP-efficient,divide-and-conquer-qsp,given_rotation,QSP-systhesis}. One approach involves introducing a `compressed' register of $\mathcal{O} (\log_2 K)$ qubits, with its basis states $\ket{l_j}$ corresponding to the determinants $\ket{D_j}$. We prepare first a state $\sum_{j=0}^{K-1} \sqrt{w_j} \ket{l_j}$ in such a compressed register. This can be achieved with $\mathcal{O}(K)$ gates \cite{QSP-systhesis}.  By implementing an isometry transformation, the basis states of the compressed register can then be mapped back to the corresponding encoded Slater determinants on $N$ qubits:
\begin{equation}
\label{compres}
    \ket{l_j} \mapsto \ket{D_j}, \quad \text{for all }  j= 0, 1,2, \cdots ,K-1.
\end{equation}
This isometry transformation can be done through the select unitary method introduced in Ref.~\cite{QSP-iso} or, more generally, through the method introduced in Ref.~\cite{QSP-QROM}. Consequently, the initial states $\ket{\Phi(\bm{w})}$ in Eq.~\eqref{wfield} can be prepared as ``generalized" Slater determinants, denoted as $\ket{D'_j} = \ket{D_j} \otimes \ket{a_j}$, where $\ket{D_j}$ corresponds to the desired reference states  and $\ket{a_j}$ can be any orthogonal state in the ancilla-qubit space:
\begin{align}
\label{QSP}
    \ket{\Phi(\bm{w})} =\sum_j \sqrt{w_j} \ket{D_j'} =\sum_j \sqrt{w_j} \ket{D_j} \otimes \ket{a_j}\,.
\end{align}

To illustrate the state preparation procedure we have just described, we utilize as an example the H$_2$ molecular system in the STO-3G basis.
In the JW mapping with an interleaved order, the desired reference states (with $S_z$=0) are represented as
$\ket{1100}, \ket{1001}, \ket{0110}$ and $\ket{0011}$.
Our objective is to prepare a weighted superposition of reference states denoted as
$\sqrt{w_0} \ket{1100}\otimes\ket{a_0}  +  \sqrt{w_1} \ket{1001}\otimes\ket{a_1} + \sqrt{w_2} \ket{0110}\otimes\ket{a_2}  +\sqrt{w_3} \ket{0011}\otimes\ket{a_3}$. To achieve this, we first create a compressed state in 2 qubits as
\begin{align*}
\sum_{j=0}^{3} \sqrt{w_j} \ket{l_j}, \quad {\rm where} \quad\ket{l_j} = \ket{00}, \ket{10}, \ket{01}, \ket{11}\,.
\end{align*}
Subsequently, we select an isometry unitary to map this state to a 6-qubit space comprising 4 working qubits and 2 ancilla qubits (see below Sec.~\ref{circuit} for more details). It is important to note that the quantum state preparation procedure is not unique and depends on the user's selection of reference states.

\subsection{Unitary ansatz: unitary coupled-cluster generalized singles and doubles}

Numerous methodologies already exist for formulating the ansatz circuit for ground and excited states. These include but are not limited to the unitary coupled cluster with single and double excitations (UCCSD) \cite{ucc-1}, the generalized unitary couple cluster (GUCC) \cite{uccgsd,Nooijen_paper}, the k-unitary pair coupled-cluster generalized singles and doubles (k-UpCCGSD) \cite{uccgsd}, the qubit CC \cite{Ryabinkin2019}, the adaptive ansatz \cite{adaptive-vqe-1}, as well as the hardware-efficient ansatz \cite{HEA}. Unlike unitary coupled cluster, which discriminates between occupied and unoccupied orbitals, GUCC treats all orbitals uniformly. This uniformity makes GUCC couple excited determinants together, thereby elevating its precision compared to the traditional UCC method. The increased expressiveness of GUCC thus results in enhanced accuracy when simulating many-body quantum systems. Given these significant advantages —particularly in scenarios requiring high precision and expressibility— we have employed GUCC as our ansatz. We now offer a brief recap of the GUCC approach.

Inspired by the standard unitary coupled cluster theory, the generalized unitary coupled cluster ansatz can be expressed as:
\begin{equation}
    \ket{\psi} = e ^{T-T^\dagger} \ket{\psi_0}\,,
\end{equation}
where $T=\sum_i T_i $ and
\begin{equation}
\begin{aligned}
T_1 &= \sum_{pq} t_{pq} a_p^\dagger a_q \,, \\
T_2 &= \sum_{pqrs} t_{pqrs} a_p^\dagger a_q^\dagger a_r a_s\,.
\end{aligned}
\end{equation}
Higher excitations can be written in a similar way.
Here, the term `generalized' means that the excitation terms do not distinguish between occupied and unoccupied spin-orbitals (i.e., the indices $p$, $q$, $r$, and $s$  denote any spin-orbitals, regardless of their occupation).
This feature allows GUCC to couple excited determinants effectively, especially at lower truncations \cite{ucc_qcview};  however, it is worth noting that including more and more transitions can result in increased computational complexity.
The unitary coupled cluster generalized singles and doubles (UCCGSD) ansatz, first mentioned in the work by Nooijen \cite{Nooijen_paper}, considers only generalized singles ($T_1$) and doubles ($T_2$) cluster operators  and finds widespread usage in the quantum computing literature \cite{PRA_Troyer,uccgsd,uccgsd-cqc,uccgsd-spin-adapt,ucc_qcview}. In this work, we utilize the UCCGSD approach and specifically concentrate on excitations that preserve the $S_z$ value instead of considering all generalized excitations.
Similar methods follow the same practice ~\cite{uccgsd-cqc,uccgsd-spin-adapt}.
To enable the practical deployment of cluster operator exponentials on quantum computers, we resort to finite step Trotterization. In this study, we used a single Trotter step, which has been shown to provide precise results \cite{ucc-Strategies,IBM-PH}.  Although the ansatz we employ is not a novel concept, our targeted approach enables us to reduce the number of variational parameters and gate counts relative to the standard UCCGSD while maintaining the excited states within the desired spin sector.

\subsection{Algorithm and quantum circuit}
\label{circuit}

To target the $K$ lowest-energy eigenstates (including the ground state) of a Hamiltonian $\hat H$ simultaneously, we need to use $\mathcal{O}(\log_2 K)$ ancilla qubits. The purified SSVQE algorithm runs as follows:
\begin{enumerate}
    \item Prepare the state $|\Phi\rangle = \sum_{j=0}^{K-1} \sqrt{w_j} |\phi_j\rangle \otimes |a_j\rangle$ with the procedure described in Sec.~\ref{sec:fixing_particle_number}.
    \item Find the parameter set $\boldsymbol{\theta}^{*}$ that minimizes the ensemble energy $\mathcal{L}_{\bd{w}}(\boldsymbol{\theta})$ \eqref{pssveqcost}.
    \item Extract the eigenstates by applying the unitary transformation $|\varepsilon_j\rangle = U(\boldsymbol{\theta}^{*})|\phi_j\rangle$ and compute the corresponding energies.
\end{enumerate}

Now we describe each step in detail. First, take $K$ initial reference (orthogonal) states $\{|\phi_j\rangle\}$. Once we ensure they have the right symmetries, it is convenient to use knowledge of the problem for choosing them. For instance, in the next section, we select Hartree-Fock states. Those are guaranteed to be close to the actual solutions in some parameter regime. As in any other minimization problem, a clever choice at this stage will impact the convergence rate. This is the only step where operations on the ancilla qubits are needed.

Second, minimizing the ensemble energy is done using standard optimization algorithms such as gradient descent. For VQE implementations, the ensemble energy is extracted by measuring the Hamiltonian $\hat{H}$ in the working qubit register. Only one measurement of the Hamiltonian expectation value is needed, in stark contrast to other methods that require one Hamiltonian measurement per targeted state \cite{PhysRevResearch.1.033062}. Finally, each eigenstate is prepared in the quantum device as $|\varepsilon_j\rangle = U(\boldsymbol{\theta}^{*})|\phi_j\rangle$. Hence, extracting $K$ eigenenergies implies measuring $\hat{H}$ separately for each eigenstate  $|\varepsilon_j\rangle$.

\begin{figure}[t]
\centering
\includegraphics[width=8.5cm]{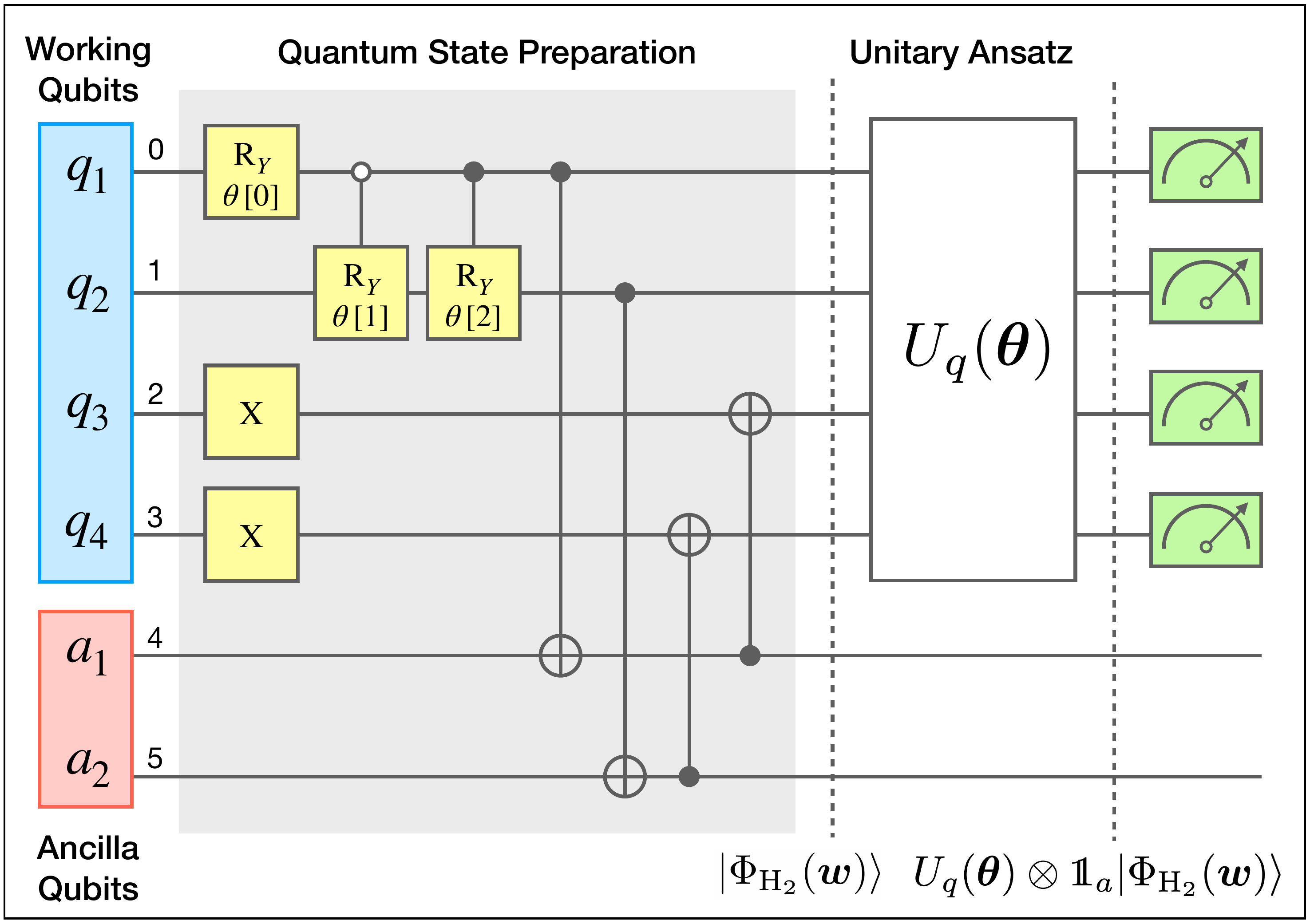}
 \caption{Sketch of the architecture of the purified SSVQE algorithm implemented to compute  the full spectrum of the dissociation of H$_2$. The first step is quantum state preparation, designed to prepare the desired state  $\ket{\Phi_{{\rm H}_2}(\bm{w})}$.  After the application of the unitary operation $U_{q}(\boldsymbol{\theta})$, a new superposition ${U_{q}(\boldsymbol{\theta}) \otimes \openone_a \ket{\Phi_{{\rm H}_2}(\bm{w})}}$ is obtained. Eventually, the unitary is optimized and the eigenpairs are extracted.}
\label{Fig:q_circ}
\end{figure}

A sketch of the algorithm and the quantum circuits are depicted in Figure~\ref{Fig:q_circ}. In the first step, the quan\-tum state preparation procedure operates on both working qubits ($q$) and ancilla qubits ($a$) to prepare the state described by Eq.~\ref{QSP}. For the case of the H$_2$ molecule in the STO-3G basis, the $R_Y$ and control $R_Y$ act on the first two qubits to prepare the `compressed' states. They contain the parameters that make up the different weights in the ensemble energy. Then, by applying two X and CNOT gates, these compressed states are mapped to obtain the linear superposition of working and ancilla states. The final result is
\begin{align}
\ket{\Phi_{{\rm H}_2}(\bm{w})} &= \sqrt{w_0} \ket{1100}_{q}\ket{00}_{a}  +
    \sqrt{w_1} \ket{1001}_{q}\ket{01}_{a} \nonumber \\ &+ \sqrt{w_2} \ket{0110}_{q}\ket{10}_{a}  +\sqrt{w_3} \ket{0011}_{q}\ket{11}_{a}  \label{QSP1} \,.
    \end{align}
    A step-by-step explanation of the quantum preparation of this state can be found in Appendix \ref{appa}.
Subsequently, the parameterized ansatz circuit $U_{q}(\boldsymbol{\theta})$ is applied only on the working-qubits space to obtain the ensemble energy $\mathcal{L}_{\bd{w}}(\bd{\theta})$. Once the optimal parameters $\boldsymbol{\theta}^{*}$ are obtained, one can extract the eigenstates by applying the unitary transformation to the initial set of working states (e.g., $|\varepsilon_0\rangle = U(\boldsymbol{\theta}^{*})|1100\rangle$) and compute the corresponding energies as the expectation values  $\langle \varepsilon_j|\hat{H}|\varepsilon_j\rangle = \varepsilon_j$.

As already pointed out above, when discussing the number of samples needed for achieving certain accuracy in the computation of the weighted energy $\sum_j w_j E_j(\bd{\theta})$, weighted SSVQE can make use of importance sampling. Under this scheme, measuring the ensemble energy with a given expected error $\varepsilon$ in our approach and weighted-random SSVQE requires the same number of samples. However, in practice, the outcomes of both methods differ greatly in the presence of noise. While in purified SSVQE the same circuit is prepared several times for each measurement, weighted-random SSVQE prepares $K$ different circuits with a given probability distribution. This has important consequences for error mitigation in noisy near-term quantum computers \cite{cai_quantum_2023}. On one hand, our implementation can be seen as an extension of standard VQE which allows the straight application of error mitigation schemes \cite{cai_quantum_2023,temme_error_2017,van_den_berg_probabilistic_2023,lolur_reference-state_2023,kurita_synergetic_2023,rogers_error_2023}. On the other hand, the known error mitigation techniques \cite{cai_quantum_2023} are not tailored for a sophisticated sampling of circuits like the importance sampling of weighted SSVQE. To be precise, the main error mitigation techniques \cite{temme_error_2017,cai_quantum_2023} rely on characterizing errors in the circuit preparing the state to be measured, which is challenging even for simple circuits \cite{van_den_berg_probabilistic_2023}.

\subsection{Numerical experiments}
\label{sec5}

 We present the numerical results validating the effectiveness of the algorithm for the dissociation of H$_2$, LiH, and the linear H$_4$. All calculations were performed using the minimal STO-3G basis set.
The qubit Hamiltonian was constructed using the OpenFermion \cite{Openfermion} and OpenFermion-PySCF \cite{Openfermion,pyscf} plugins. Quantum simulations were conducted using PennyLane \cite{pennylane} and Qiskit \cite{Qiskit}. The optimization of circuit parameters utilized the Adam \cite{adam} algorithm. For the calculation of the first $K$ eigenenergies the vector $\bd{w}$ utilized in all simulations is chosen as $(K, K-1, \cdots, 1)$ and then normalized to 1. To ensure accurate results, the convergence criterion for the ensemble energy threshold was set to $10^{-9}$ Hartree. For complete access to the code and settings utilized in this work, please visit the GitHub repository \cite{github_code}.

\begin{figure}[t]
\centering
\includegraphics[width=8.5cm]{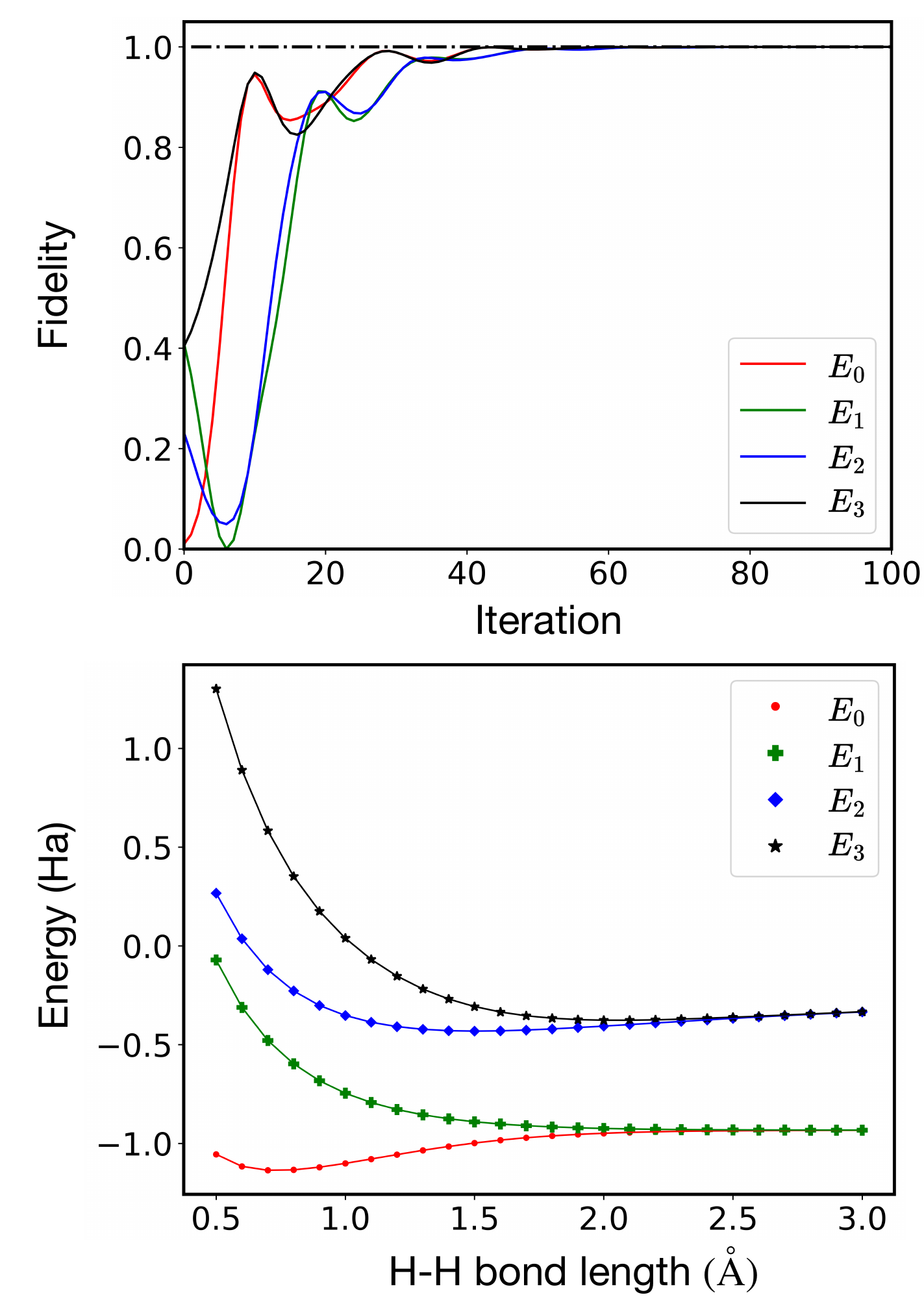}
 \caption{Computation of the spectrum of the dissociation of H$_2$ using purified SSVQE. The upper panel illustrates the evolution of the state fidelities during the optimization process for a fixed bond length. The lower panel presents the comparison between the computed electronic eigenenergies and the exact diagonalization results (solid color line) along the dissociation path.
 }
\label{Fig:h2_pes}
\end{figure}

\begin{figure*}[ht!]
\includegraphics[width=0.85 \textwidth]{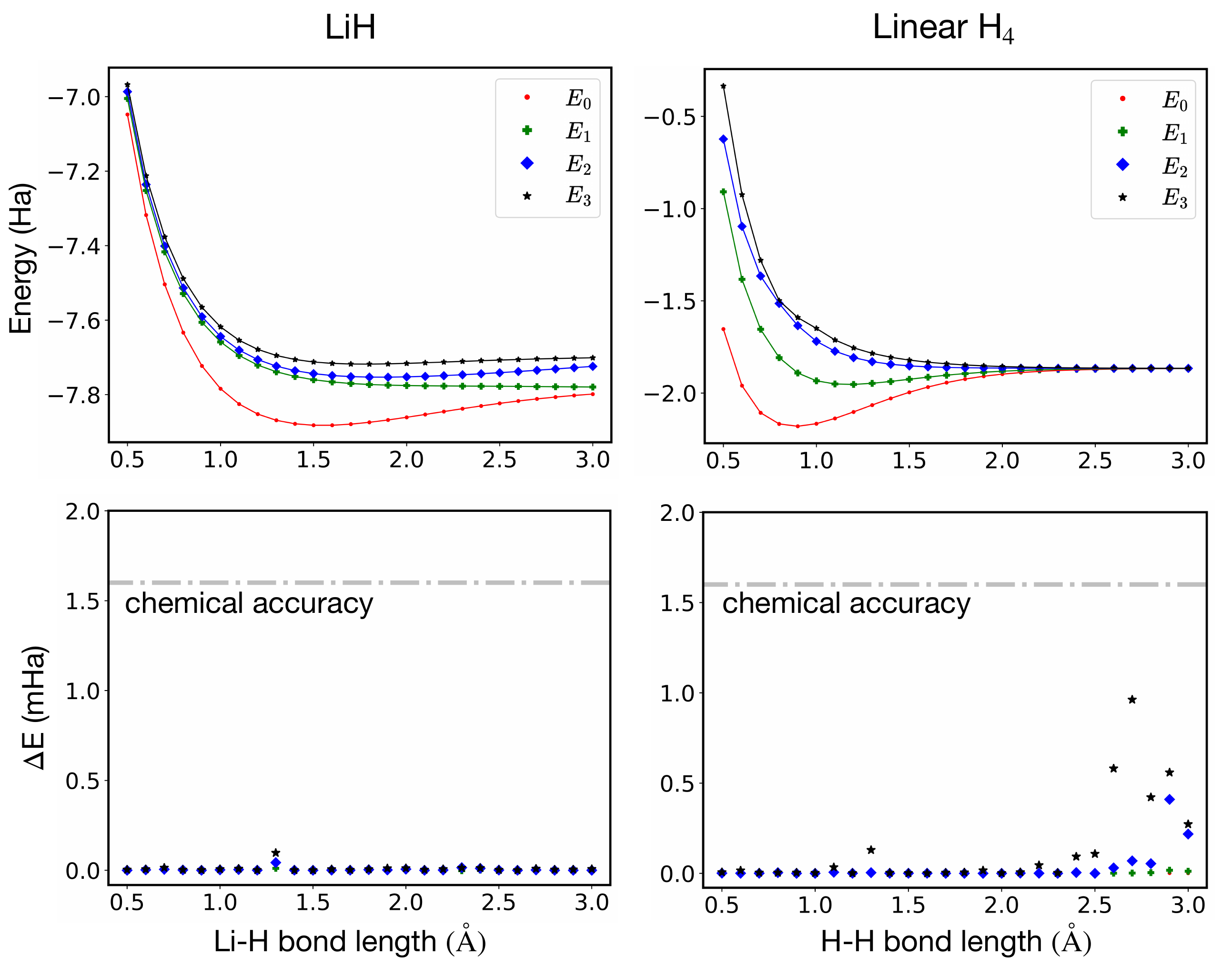}
\caption{Ground and excited-state energies of (a) LiH and (b) Linear H$_4$. The lines correspond to the results obtained using the exact diagonalization (ED) method, while the dots represent our calculation using our purified SSVQE. Energy differences ($\Delta E = E_{\rm predicted} -E_{\rm ED}$), measured in mHartree, are presented for (c) LiH and (d) Linear H$_4$. The dash-dot silver line represents the level of chemical accuracy.}
\label{lih-h4}
\end{figure*}

For the H$_2$ molecule, which is described by 2 spatial orbitals in the STO-3G basis set, we required 4 qubits for storing the state information using JW encoding. We calculated the Qubit Hamiltonian for bond lengths ranging from 0.5 to 3.0 {\AA} with intervals of 0.1 {\AA}. For the H$_2$ molecule, we performed computations for the complete eigenspectrum within the sector $S_z = 0$. Figure~\ref{Fig:h2_pes} showcases the fidelity $|\braket{\psi_{\text{computed}}}{\psi_{\text{ED}}}|^2$ of each individual state through the optimization process. This result gives a first demonstration of the effectiveness of this approach to estimate, simultaneously, the excited states. The potential energy curves are also shown in  Figure~\ref{Fig:h2_pes}. The computed results for all excited-state energies are in complete agreement with the exact diagonalization (ED) method. From the chemical perspective, it is understandable why the UCCGSD ansatz was able to generate, in this case, precise outco\-mes when calculating the electronic states of the hydrogen molecule in a minimal basis set. This is due to the fact that the UCCGSD ansatz considers all electronic excitations of the two electrons in the system, allowing it to exactly capture all the possible configurations of H$_2$.
This ability of UCCGSD to expand the full Hilbert space of 2 electrons (or 2 holes) was also observed for the classical computation of the spinless Fermi-Hubbard model with five sites \cite{PhysRevLett.129.066401}.

\begin{figure*}[t]
\centering
\includegraphics[width=1.0 \textwidth]{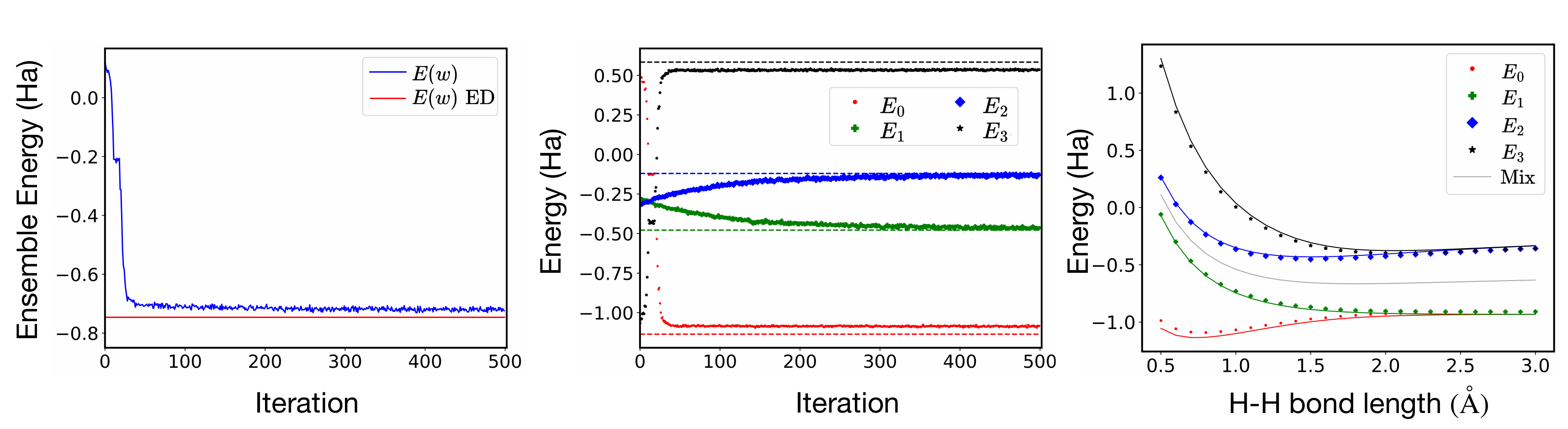}
\caption{Left: Optimization of the ensemble energy for H-H length=0.7 \AA, plotted against the exact ensemble energy (solid red line). Middle: The convergence of individual states during the optimization against the exact value (color dashed line).
Right: Potential energy curves and computational errors of H$_2$ obtained from calculations in the presence of noise. The solid color lines depict the eigenenergies and the equiensemble energy obtained using exact diagonalization. The grey line denotes $\Tr(H)/4$ which is the ensemble energy of a totally mixed state.
}
\label{Fig:h2_noise}
\end{figure*}

In the case of the LiH molecule, which is described by 6 spatial orbitals (12 spin-orbitals) within the STO-3G basis set, we adopt the frozen core approximation to streamline the computational process. We assume that the 1s orbital of Li is always doubly occupied and freeze it using the frozen core approximation. Accordingly, we only consider the remaining 10 spin-orbitals, which require 10 qubits using the JW encoding scheme. The potential energy curves and corresponding errors are shown in Figure~\ref{lih-h4} (a) and (c), respectively, where the ground state ($E_0$) and the first three excited states ($E_1, E_2$ and $E_3$) are calculated. As expected, the calculated results for all excited states align with the exact diagonalization method  \cite{uccgsd}. Moreover, the relative errors in the calculations reach chemical accuracy (1.6 mHa).

Despite its simplicity, the linear H$_4$ system  has emer\-ged as a popular choice for investigating novel numerical approaches. With only four electrons, this system exhibits strong static electron correlations \cite{C7CP01137G}, making it an ideal platform for investigating the impacts of electron interactions and benchmarking various computational methods. The outcomes of the calculations, including the ground state and the first three excited states, are illustrated in Figure~\ref{lih-h4} (b) and (d). It is noteworthy that, based on Figure~\ref{lih-h4} (d), higher excited states exhibit larger errors than the lower ones.  This behavior can be ascribed to the enhanced complexity inherent to higher excited states, which subsequently imposes further challenges when precisely computing their energies. Also, the smaller value of the hyperparameters $w_j$ for higher states, which are less attended during optimization, leads to an increase in error. As the bond length increases, the states demonstrate a trend toward degeneracy. This phenomenon arises from the weakening of interatomic interactions and the gradual transformation of molecular energy levels into atomic ones. It is important to note that the final optimization result is influenced by the selection of reference states and the initial values of variational parameters. If the chosen reference states accurately capture the characteristics of the desired $K$'th excited state, the convergence to the global minimum can be achieved with fewer iterations.

\subsection{Robustness against errors in quantum hardware}
\label{noise}

After having established the efficacy of the proposed framework in classical noiseless simulation, we proceed now to introduce noise into the simulations. To solely investigate the pure-quantum aspects of the system, we disregard the readout error.
The noise model employed is based on the realistic error statistics of the ibmq$\_$manila device (see Appendix~\ref{calibration} for the comprehensive information of the device).  To minimize the required number of qubits and gate operations and optimize resource utilization, we employ Checksum encoding \cite{check_code_mapping}, instead of JW mapping. By utilizing this encoding technique, we are able to save 2 qubits for the H$_2$ molecule. Furthermore, we only select effective excitations (only include two independent double excitations), which reduces both the circuit depth and the number of parameters, leading to more practical computations \cite{Higgott2019variationalquantum,QC-iontrap}.
In the presence of noise, we utilize the simultaneous perturbation stochastic approximation (SPSA) optimization algorithm, which is effective in situations involving measurement uncertainty on quantum computation (when finding a minimum) \cite{HEA}. We take $10^4$ measurement shots in each iteration to estimate the expected value of the energy.

Figure~\ref{Fig:h2_noise} shows that both the ensemble energy and individual energies exhibit convergence in the noisy simulation. We observe this trend: the ensemble energy oscillates about a stabilized average value which lies higher than the exact result. Interestingly, while the first and second excited state reaches the exact value, the ground, and third excited state does not. The variational algorithm's capability in obtaining `correct' parameters \cite{vqe-theory} makes it possible for ensemble energies to stabilize at a specific level. Yet, the effect of noise (in particular, the depolarization errors prevalent in NISQ devices) may not be evenly dispersed across all individual energies. It can be observed in Figure~\ref{Fig:h2_noise} that the ground and the first-excited energies obtained under noisy conditions are higher than the ED results. Conversely, the second and third excited state results are lower than ED, suggesting that they exhibit characteristics of a totally mixed ensemble (see gray line). We can then predict the trend of the noisy energies according to their position with respect to the totally mixed state $\Tr(H)/4$. This fact could be exploited in the development of error mitigation techniques tailored to ensemble energies. Thus, purified SSVQE shows intrinsic resilience to noise, which can be potentially improved through error mitigation.

\section{The role of the weights}
\label{secw}
The ensemble variational principle \eqref{costSSVQE} ensures that the (exact) solution to the problem presented in Eq.~\eqref{QPVQEcost} for minimizing the ensemble energy is independent of the specific choice of $\bd{w}$. In practical scenarios, however, as the circuit ansatz cannot express the full unitary, the choice of the weights influences both the search landscape and the convergence rate of the iterative update of the circuit parameters. In particular, when employing gradient-based optimization methods, it is expected that the quality of the individual energies and states depends significantly on $\bd{w}$. Moreover, specific conditions can lead to cost-function landscapes with exponentially diminishing gradient magnitudes (i.e., barren plateaus). This section therefore aims at elucidating the impact of the weight selection on the algorithm's performance.

\subsection{The predictive power of the ensemble variational principle}
\label{sec:V-A}

As mentioned above, the exact minimal ensemble energy $E_{\bm{w}}$ is always realized by the ensemble $\rho_{\bm{w}}$ of the exact eigenstates. One can therefore rely on the ensemble variational principle to extract from $\rho_{\bm{w}}$ the individual eigenenergies and eigenstates.
The only exception is that, if two weights are the same, e.g., $w_j = w_{j+1}$, then any resolution between the two energy levels $E_j$ and $E_{j+1}$ is lost within the framework of ensemble minimization. To further distinguish between the two energy levels, an additional subspace diagonalization step would be needed.

In practice, however, the ensemble energy almost always retains a finite error, due to either a set error threshold for convergence, or the limitations of the circuit ans\"atze. Consequently the eigenenergies are also not exact, and their errors could potentially be magnitudes larger than that of the ensemble energy, due to error cancellations. In that sense, it is \textit{a priori} not clear whether the use of the ensemble variational principle yields reliable predictions of quantities of interest, such as the energies and other properties of the eigenstates, when the ensemble energy is not \textit{exactly} converged. In Ref.~\cite{ding2024ground} some of the authors proved that the ensemble variational principle is indeed equipped with such a predictive power. To be more specific, it was found and proven that the error of any quantity $Q$ of interest is bounded as \cite{ding2024ground}:
\begin{equation}
    d_{-}^{(Q)} ({\bd{w}},\bd{E}) \Delta E_{\bd{w}} \le \Delta Q \le d_{+}^{(Q)} (\bm{w},\bm{E}) \Delta E_{\bd{w}}\,,
\label{linear-bounds}
\end{equation}
where $d_{\pm}^{(Q)} ({\bd{w}},\bd{E})$ are weight- and potentially spectrum-dependent prefactors. While $d_+^{(Q)} ({\bd{w}},\bd{E})$ is always positive, $d_-^{(Q)} ({\bd{w}},\bd{E})$ may vanish or be negative depending on $Q$. Eq.~\eqref{linear-bounds} specifies the possible range of error of the quantity $Q$, linearly related to the error of the ensemble energy. As $\Delta E_{\bm{w}}$ decreases, the maximal error that can incur in the quantity $Q$ also reduces.

Given the linear bounds in Eq.~\eqref{linear-bounds}, it is therefore natural to propose an ``optimal'' weight vector $\bm{w}^{(Q)}$ for each quantity $Q$ that minimizes the prefactor  $d_{+}^{(Q)}(\bm{w},\bm{E})$ of the analytical upper bound, and thus ``maximizes'' the predictive power of the ensemble variational principle in the accuracy of $Q$. We note that $\bm{w}^{(Q)}$ is only optimal in the sense that for a $\boldsymbol{w}$-independent error in the ensemble energy $\Delta E_{\bm{w}}$, the upper bound of $\Delta Q$ is minimized. The weight dependence of the error $\Delta E_{\bm{w}}$ is not taken into our consideration. As we will see in the next sections, this simplification does not undermine the advantage of the ``optimal'' weights $\bm{w}^{(Q)}$. In the following, we will continue to use the quotation mark to remind the reader of the context of the term ``optimal''. For the sake of completeness and to complement our subsequent numerical analysis, we recall from Ref.~\cite{ding2024ground} the ``optimal'' weights for several quantities of interest and list them in Table \ref{GOK-table}.

\begin{table*}[ht]
\centering
    \begin{tabular}{|c|c|c|c|}
    \hline
    \rule{0pt}{0ex}\rule[-1ex]{0pt}{0pt}
    \textbf{Quantity} & \textbf{Error}  & \textbf{Lowest Upper Bound}& \textbf{Optimal} $\wb$
    \\
    \hline
    \rule{0pt}{3.5ex}\rule[-5.5ex]{0pt}{0pt}
    $\ket{\Psi_j}$ &$\Delta \Psi_j$
    & $\:k(r_-\!+\!r_+)\!+\!r_+$
    & $\wb \propto (\underbrace{r_{-}\!+\!r_{+},\ldots,r_{-}\!+\!r_{+}}_{j-1},r_{+},\underbrace{0,\ldots,0}_{D-j})$
    \\
    \hline
    \rule{0pt}{4.5ex}\rule[-3ex]{0pt}{0pt}
    $\{\ket{\Psi_j}\}_{j=0}^{K-1}$&$\sum_{j=0}^{K-1}\Delta \Psi_j$  &  $\sum_{j=1}^{K}2k(E_{j}\!-\!E_{j-1})^{-1}\Theta(K\!-\!j)$ & $\bm{w}\propto \left(\sum_{l=j+1}^{D-1}(E_{l}\!-\!E_{l-1})^{-1}\Theta(K\!-\!l)\right)_{j=0}^{D-1}$
    \\
    \hline
    \rule{0pt}{3.5ex}\rule[-5.5ex]{0pt}{0pt}
    $E_j$ & $|\Delta E_j|$  & $2j+1$ & $\bm{w}\propto(\underbrace{2,\ldots,2}_{j},1,\underbrace{0,\ldots,0}_{D-j-1})$
    \\
    \hline
    \rule{0pt}{3.5ex}\rule[-2ex]{0pt}{0pt}
    $\{E_j\}_{j=0}^{K-1}$ & $\sum_{j=0}^{K-1}|\Delta E_j|$    &  $K^2$ & $\bm{w}\propto(2K\!-\!1,2K\!-\!3,\ldots,1,0,\ldots,0)$
    \\
    \hline
    \end{tabular}
\caption{The lowest upper bounds and the corresponding optimal choice of $\bm{w}$ (normalized to 1) for various errors of the quantum states and eigenenergies \cite{ding2024ground}. We defined $r_{\pm} \equiv |E_{j}-E_{j\pm1}|^{-1}$, $D$ is the dimension of the Hilbert space, $K<D-1$, and $\Theta(x)$ denotes the Heaviside step function.
}
\label{GOK-table}
\end{table*}

\subsection{Model system and simulation setting}
We now illustrate the role of the weights in our algorithm, with the Transverse Field Ising Model (TFIM), whose Hamiltonian reads:
\begin{equation}
H = \sum_{i=1}^N a_i X_i + \sum_{1 \leq i < j \leq N} J_{ij} Z_i Z_j,
\label{eq:TFIM}
\end{equation}
where $X_i, Z_i$ denote the Pauli operators for the $i$-th spin. This model is exactly solvable \cite{pfeuty1970one}, allowing us to compare the new results obtained with our implementation with the exact ones. For the purpose of the demonstration, we select $N=4$ and focus on the first $K=3$ eigenenergies and eigenstates. We use the circuit ansatz proposed in Ref.~\cite{PhysRevA.101.032308}, with a depth of $D_{c}=18$. The circuit parameters were optimized using the Adam algorithm with a learning rate of $lr=0.01$. The initial values of the circuit parameters are randomly sampled from a uniform distribution $[0, 2\pi)$, and are fixed across different choices of weights.

Under this setting, we investigate the impact of different weight vectors $\bm{w} = (w_0,w_1,w_2) \equiv (w_0,w_1,1-w_0-w_1)$ by sampling 1000 different $(w_0, w_1)$ pairs from the convex set:
\begin{equation}
    \mathcal{P} = \left\{(w_0,w_1)\,\left|\, w_0 \!>\! w_1 \!>\! 1\!-\!w_0\!-\!w_1 \!>\! 0\right.\right\} \in \mathbb{R}^2,
\end{equation}
by focusing on two physical scenarios:
\begin{enumerate}
    \item Non-degenerate, nearly equal energy spacing: In this scenario, the targeted energy levels of the first \(K\) states of interest are nearly equally spaced.
    \item Nearly degenerate energy spacing: In this scenario, some (but not all) targeted energy levels are nearly degenerate.
\end{enumerate}

\subsection{Numerical Results}
\subsubsection{Non-degenerate, nearly equal energy spacing}
\label{TFIM1}
In the following, we consider the Hamiltonian \eqref{eq:TFIM} with $(a_1,a_2,a_3,a_4)=(0.40547, 0.48914,0.71003,0.24241)$ and $(J_{12}, J_{23}, J_{34}) = (0.90389, 0.16600, 0.76043)$ which results in the following first eigenenergies
\begin{equation}
\begin{split}
E_0&=-2.51396168, \\
E_1&=-2.26570123, \\
E_2&=-2.03866159, \\
E_3&=-1.79040113, \\
E_4&=-0.41777537.
\end{split}
\label{TFIM1-spec}
\end{equation}

\begin{figure*}[ht!]
\includegraphics[width=0.95 \textwidth]{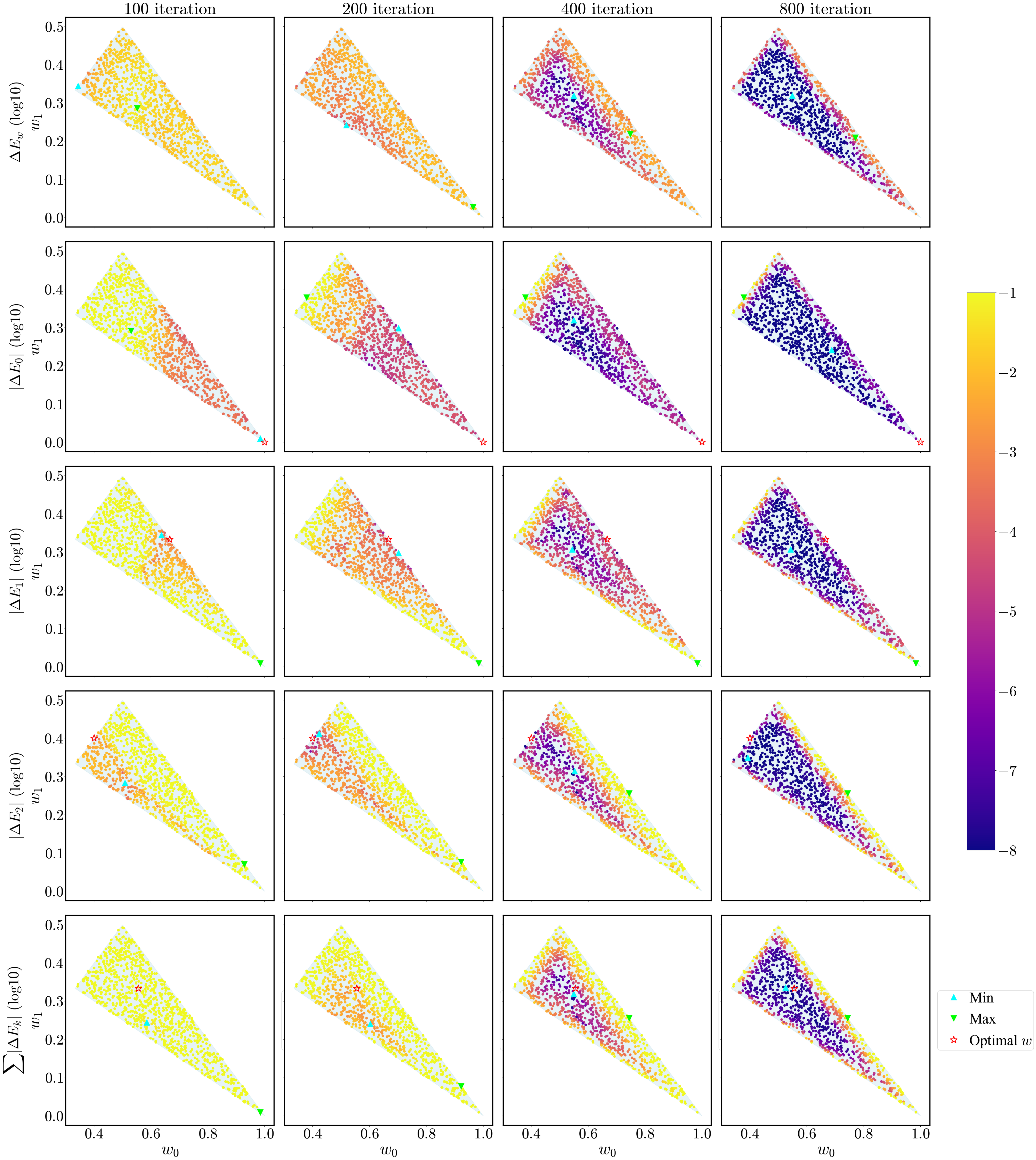}
\caption{Convergence analysis with different weights across iterations. The figure illustrates the impact of the weight selection on the convergence of energy errors (e.g., $\Delta E_{\bd{w}}$, $|\Delta E_1|$, $|\Delta E_2|$) across iterations from 100 to 800. Green (cyan) markers indicate the numerical minimum (maximum) error for the interested quantities in that iteration step and red stars represent the optimal weight selection according to Ref.~\cite{ding2024ground}.}
\label{fig:w-energies}
\end{figure*}

\begin{figure*}[ht!]
\includegraphics[width=0.95 \textwidth]{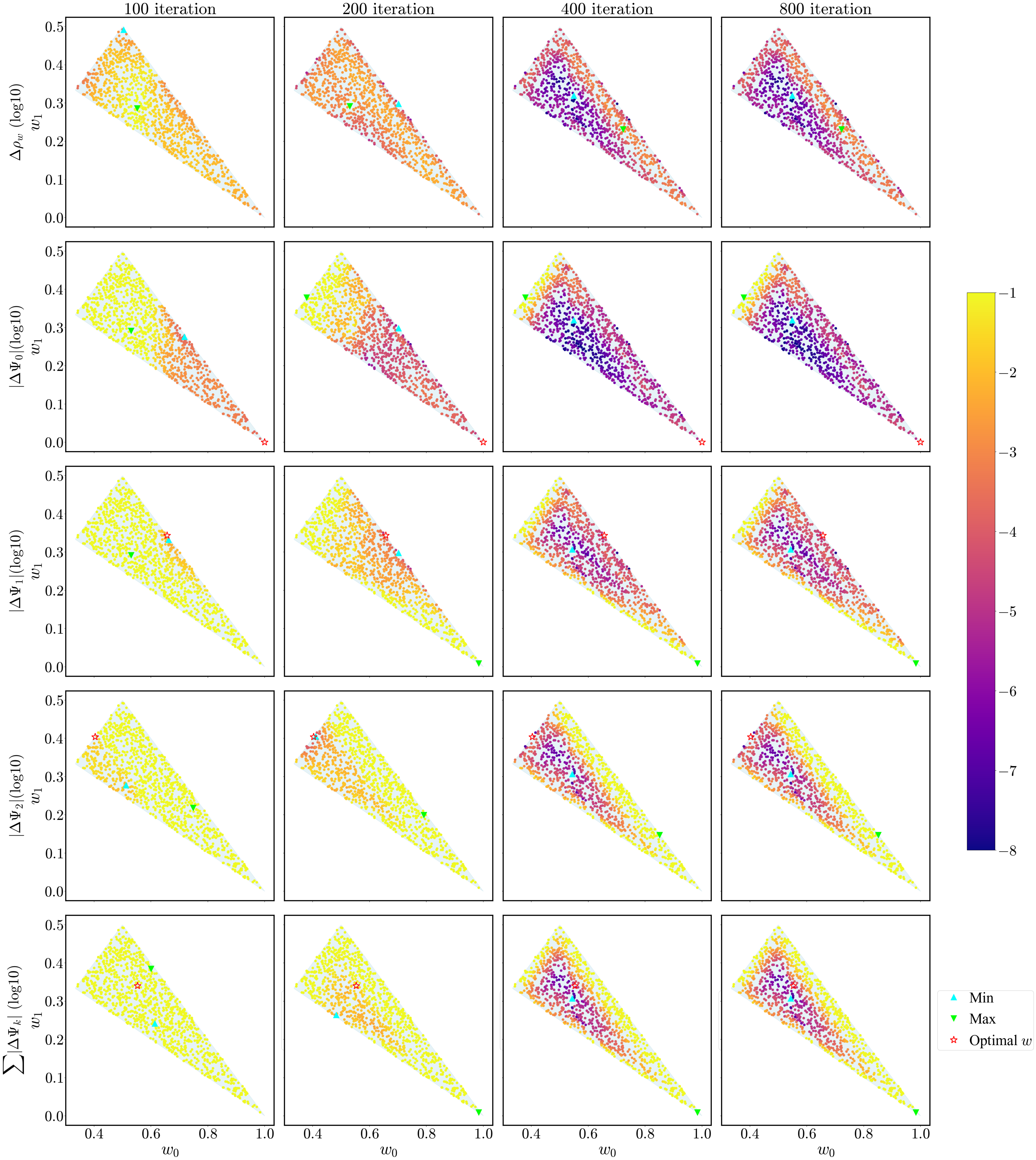}
\caption{Convergence analysis of algorithm with different weight selections across iterations. The figure illustrates the impact of the weight selection on the convergence of errors $\Delta \rho_w$ and relevant state quantities across iterations from 100 to 800. Green (cyan) markers indicate the numerical minimum (maximum) error for the interested quantities in that iteration step and red stars represent the optimal weight selection according to \cite{ding2024ground}.}
\label{fig:w-states}
\end{figure*}

Figs.~\ref{fig:w-energies} and \ref{fig:w-states} display the individual and ensemble energy and state error dependencies based on random weight selections. They show how well our purified SSVQE (or importance sampling SSVQE) performs across different iterations. Each figure is divided into several rows, each representing the convergence evolution of a specific physical quantity $Q$ (such as total energy error, individual energy error, or ensemble state error) over iterations ranging from 100 to 800. The top row displays the total energy (state) error $\Delta E_{\bd{w}}$ ($\Delta \rho_{\bd{w}}=\left\| \rho_{\text{computed}} -\rho_{\rm exact} \right\|_{\rm HS}^2 $) on a logarithmic scale. As we discussed in Section \ref{sec:V-A}, when the ensemble energy difference $\Delta E_{\bd{w}}$ converges well, the same holds for the ensemble state error $\Delta \rho_{\bd{w}}$, as well as for the errors of the individual eigenenergies $\Delta E_j$, and those of the eigenstates $\Delta \Psi_j$ measured by the Hilbert-Schmidt (HS) distance. The subsequent rows illustrate the quality of the numerical result of various individual physical quantities as a function of the number of iterations. Green (cyan) markers indicate the numerical minimum (maximum) error in that iteration step. The red stars represent a theoretical ``optimal'' weight vector of each physical quantity $Q$, which minimizes the prefactor $d_{+}^{(Q)}(\bm{w},\bm{E})$ of the analytical upper bound.

If the ansatz of the unitary is expressive enough and a large number of iterations is executed, the selection of ${\bd{w}}$ will not affect the final optimized solution \cite{PhysRevA.87.062501,wheeler2024ensemble}. This is the interpretation of the fourth column of Figs.~\ref{fig:w-energies} and \ref{fig:w-states}, where a large area of dark blue indicates that the error differences in absolute value become small across a wide range of ${\bd{w}}$. In fact, the specific choice of the weight ${\bd{w}}$ does not play a critical role in the execution of the algorithm as long as the convergence is achieved with machine precision. In early iterations, however, the situation is quite different. The first three columns of Figs.~\ref{fig:w-energies} and \ref{fig:w-states} suggest that some regions of weight selection lead to faster convergence.
Qualitatively, we observe that points near, but not exactly on, the boundary exhibit slower convergence in $\Delta E_w$, as shown in the first row of the plot. Weight vectors close to the boundary have two elements $w_j$ and $w_{j+1}$ very close to each other, and this can cause some gradients to almost vanish. This can be rationalized as the following: if $w_j \approx w_{j+1}$, then any rotation within the subspace spanned by the $j$-th and $j\!+\!1$-th eigenstate in the ensemble state would only lead to a small but non-negligible change in the ensemble energy. And since we follow a first-order gradient method, the degeneracy in $\bm w$ can lead to slow convergence.
Moreover, as explained in Ref.~\cite{bierman2022quantum}, every energy level contributes differently to the overall cost function, with lower energy levels contributing more than the higher ones (as they are multiplied by a larger weight). As the ensemble energy is minimized, the energy level contributing more to the weighted sum are prioritized. This effect can be distinctively observed in the second, third, and fourth rows of Figure~\ref{fig:w-energies}: $\Delta E_0$ converges rapidly, as indicated by the larger darker areas in each specific iteration, compared to the smaller darker area in $\Delta E_1$ and $\Delta E_2$. Once $\Delta E_0$ reaches a sufficiently low error, $\Delta E_1$ converges to a certain accuracy. This sequential convergence continues with $\Delta E_2$, which only improves after $\Delta E_0$ and $\Delta E_1$ have sufficiently decreased in error. An important point to remember is that even though the ensemble variational principle allows us to target multiple eigenstates simultaneously, the ``optimal'' weights for targeting different energy levels are also different. For example, the weight vector $\bm{w} = (1,0,0)$ (corresponding to the Rayleigh-Ritz scenario) is ``optimal'' for targeting the ground state, but does not allow us to obtain any information of the excited states via an ensemble minimization. Therefore to collectively assess the impact of the choice of the weights over all targeted eigenenergies (eigenstates), we find it most informative quantities to inspect the sum of the absolute errors of the eigenenergies (eigenstates) $\sum_j |\Delta E_j|$ ($\sum_j \Delta \Psi_j$). In the fifth row of Figs.~\ref{fig:w-energies} and \ref{fig:w-states}, we observe that the numerical optimal weights (marked by the points with the smallest errors after a given number of iterations) is approaching the theoretical ``optimal'' weights defined in Table \ref{GOK-table}, and in the regions close to the ``optimal'' weights our approach typically converges faster. In fact, the red stars appear in polytope $\mathcal{P}$ where rapid convergence occurs, suggesting that an appropriate selection of ${\bd{w}}$ can enable the algorithm to reach the desired accuracy with fewer number of iterations.

\subsubsection{Nearly degenerate energy spacing}
\label{TFIM2}

\begin{figure*}[htb]
\includegraphics[width=0.95 \textwidth]{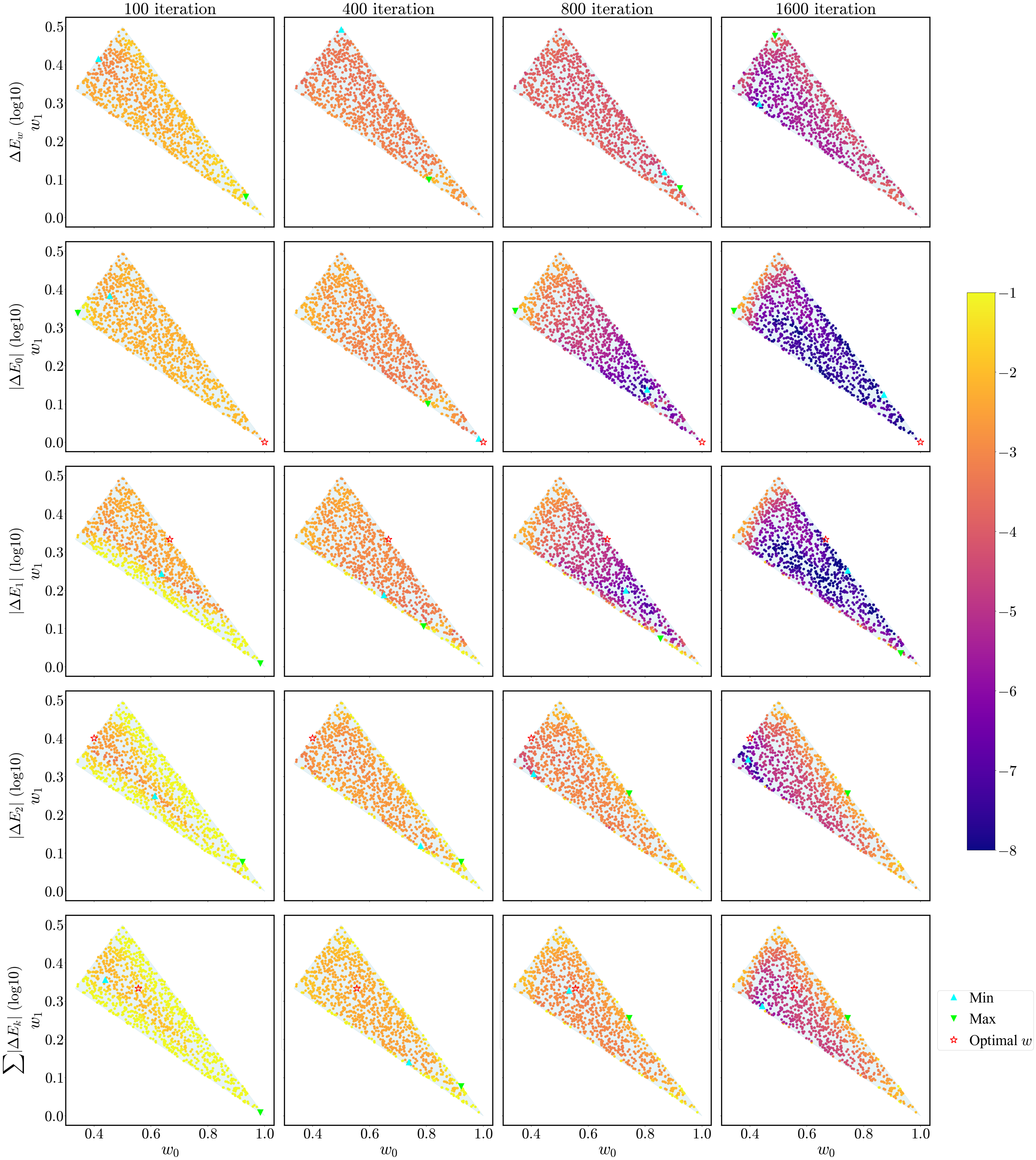}
\caption{Convergence analysis with different weights across iterations. The figure illustrates the impact of the weight selection on the convergence of energy errors (e.g., $\Delta E_{\bd{w}}$, $|\Delta E_1|$, $|\Delta E_2|$) across iterations from 100 to 1600. Green (cyan) markers indicate the numerical minimum (maximum) error for the interested quantities in that iteration step, red stars represent the optimal weight selection according to Ref.~\cite{ding2024ground}.}
\label{fig:w-energies-2}
\end{figure*}

\begin{figure*}[htb]
\includegraphics[width=0.95 \textwidth]{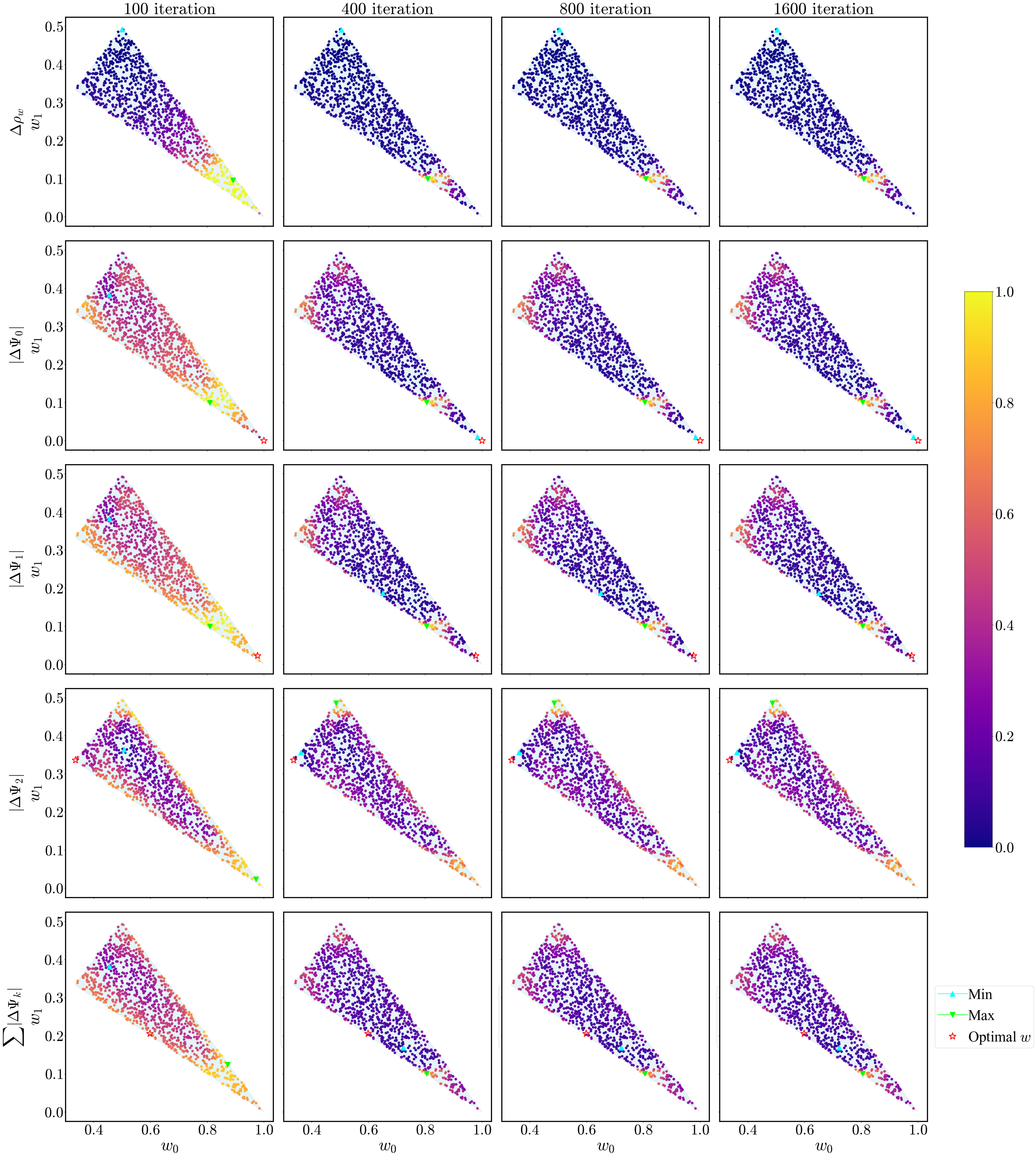}
\caption{Convergence analysis of algorithm with different weight selections across iterations for near-degenerated TFIM system. The figure illustrates the impact of the weight selection on the convergence of errors $\Delta \rho_w$ and relevant state quantities across iterations from 100 to 1600. Green (cyan) markers indicate the numerical minimum (maximum) error for the interested quantities in that iteration step, red stars represent the optimal weight selection according to
\cite{ding2024ground}.}
\label{fig:w-states-2}
\end{figure*}

The prefactors in Eq.~\eqref{linear-bounds} can also depend on the energy spectrum $\bm E$ for some quantities $Q$. To thoroughly investigate how the choice of the weights could impact the search landscape and convergence rate, we extend our study to the same type of Hamiltonian \eqref{eq:TFIM} but with a nearly degenerated spectrum. In order to achieve this, the coefficients for the transverse field terms are chosen as $(a_1,a_2,a_3,a_4)=(0.01749, 0.89157, 0.28486, 0.29898)$,
and $(J_{12}, J_{23}, J_{34}) = (0.79203, 0.32447, 0.86471)$ which results in the following first few eigenenergies
\begin{equation}
\begin{split}
E_0&=-2.39891268, \\
E_1&=-2.38855921, \\
E_2&=-1.95749440, \\
E_3&=-1.94714093,  \\
E_4&=-0.49529575.
\end{split}
\label{TFIM2-spec}
\end{equation}
Figs.~\ref{fig:w-energies-2} and \ref{fig:w-states-2} illustrate the convergence behavior and the dependencies of energy and state errors. Not surprisingly, the near-degeneracy of the energies makes the optimization process more difficult, as it is more intricate to distinguish between energy levels, leading to slower convergence rates.
This is evidenced by the slower reductions in $\Delta E_{\bd{w}}$ and $\Delta \rho_{\bd{w}}$ in Figs.~\ref{fig:w-energies-2} and \ref{fig:w-states-2}.
In contrast to the previous cases, where many $\wb$-points converged (became dark blue) within 800 iterations, we had to run up to 1600 iterations in the nearly degenerate case to reach comparable accuracies. Yet achieving convergence is still challenging and incomplete in some $\wb$-regions.
To highlight the difference and the gradual convergence process, we display the infidelity ($\Delta \Psi_j $) in Fig.~\ref{fig:w-states-2} in a linear scale.
In Figure, we observe that the difference between the second and third row of Figs.~\ref{fig:w-energies-2} and \ref{fig:w-states-2} are quite small compared to Figs.~\ref{fig:w-energies} and \ref{fig:w-states}. This is expected, as the energy difference between $E_0$ and $E_1$ is much smaller in the current Hamiltonian, making it more difficult to differentiate the two states. In contrast to Fig.~\ref{fig:w-energies}, where we observe clear sequential convergence trends (with $E_0$ converging faster than $E_1$ and so on), the trends in Fig.~\ref{fig:w-energies-2} are now less significant. Nonetheless, in the third and fourth columns of Fig.~\ref{fig:w-energies-2}, the errors of $E_0$ and $E_1$ are much lower than that of $E_2$. Combining Fig.~\ref{fig:w-energies} and \ref{fig:w-energies-2}, we observe that when the gap $E_{j+1}\!-\!E_j$ is considerable, $E_j$ converges to higher accuracy faster than $E_{j+1}$. In contrast, when the gap $E_{j+1}\!-\!E_j$ is almost vanishing, the errors of $E_j$ and $E_{j+1}$ become similar.

The errors of the energies and quantum states, in this case, are less sensitive to the choice of the weights, compared to nearly equal energy spacing (see Figs.~\ref{fig:w-energies} and \ref{fig:w-states}). This can also be related to the fact that our cost function, the ensemble energy, is the weighted sum of all targeted eigenenergies. In the nearly degenerate case, we observe that the numerical gradients become much smaller (and the landscape of the ensemble energy becomes more flat). This is why regardless of the choice of the weights (equally spaced or nearly degenerate), finding the optimal circuit parameters by following the vanishing gradients remains extremely difficult.
Another major difference caused by the nearly degenerate spectrum concerns the ``optimal'' weight vector for targeting an individual eigenstate. Recall from Sec.~\ref{sec:V-A} that, in contrast to the ``optimal'' weights for $\Delta E_j$ which are independent of the energy spectrum $\bd{E}$, the ``optimal'' weights for $\Delta \Psi_j$ depend on the corresponding energy gaps. Since $E_0$ and $E_1$ are nearly degenerate, with $|E_1 \!\,-\!\, E_2|^{-1} \ll |E_1 \!\,-\!\, E_0|^{-1}$,
according to Table \ref{GOK-table}, the optimal weight selection for $\Delta \Psi_0$ and $\Delta \Psi_1$ are almost identical. Consequently, we see a similar convergence pattern comparing the second and third rows of Fig.~\ref{fig:w-states-2}. 

\section{Conclusions and outlook}

In this work, we have elaborated on fundamental aspects of the weighted SSVQE.
Explicitly designed to prepare eigen\-states \textit{simultaneously}, we have presented a purified version of the algorithm in which, instead of computing $K$ eigenstates successively, we prepare and manipulate only one state that stores all the relevant information. This simultaneous preparation of eigenstates on a quantum device offers several crucial advantages. First of all, an optimal operator averaging is achieved because sampling that pure state is equivalent to improving SSVQE by importance sampling. This key feature makes our implementation a resource-efficient version of the SSVQE algorithm when it comes to computing the cost function and its gradients. Next, we have conducted quantum simulations for several systems. The computed eigenenergies and eigenstates are in excellent agreement with results from exact diagonalization.

From a theoretical point of view, the variational principle used in SSVQE is agnostic to the choice of the weights (as long as they are distinct). However, in practical scenarios, where the ensemble energy is only minimized to a set error threshold using a gradient method, or where the circuit unitary can not exactly diagonalize the Hamiltonian, the errors in the predicted energies and states, as well as the convergence rate, become dependent on the chosen weights $\bd{w}$. To study this rather intricate dependence, we conducted an extensive study on a model system with both non-degenerate and nearly degenerate spectra. We found that the convergence rates of the energy (and states) are strongly influenced by the numerical choice of the weights, particularly in non-degenerate systems and at the early stages of the simulations. While universal conclusions are difficult to draw due to dependencies on the spectrum structure, unitary ansatz, optimization algorithm, and initial point selection, we have observed and identified that certain regions of weights may contribute to faster convergence.

Our results encourage further explorations in several directions. For instance, the algorithm (either purified SSVQE or importance sampling SSVQE) can be implemented in combination with a wide range of existing numerical methods. Just to give an example, a neural network can be trained to learn the parameters of a quantum circuit preparing thus not only a chemical ground state as in Refs.~\cite{ceroni2023generating, Shang2023} but also various eigenstates.  Similarly, since the parallelization technique we have proposed is indifferent to the quantum statistics of the underlying problem, the algorithm can also potentially be applied to study vibrational spectra which could have a significant impact on the long-sought full quantum treatment of Fermi-Bose mixtures \cite{Fregoni,PhysRevB.107.024305,BasovAsenjo}.

\begin{acknowledgments}
We gratefully acknowledge financial support from the Munich Quantum Valley, which is supported by the Bavarian state government with funds from the Hightech Agenda Bayern Plus, the Deutsche Forschungsgemeinschaft (Grant SCHI 1476/1-1), the Munich Center for Quantum Science and Technology  (C-L.H, L.D and C.S.), and the European Union’s Horizon Europe Re\-search and Innovation program  un\-der the Marie Skło\-dowska-Curie grant agreement n°101065295 (C.L.B.-R.). Views and opinions expressed are however those of the authors only and do not necessarily reflect those of the European Union or the European Research Executive Agency. L.C. gratefully acknowledges funding by the U.S. ARO Grant No. W911NF-21-1-0007. All statements of fact, opinion, or conclusions contained herein are those of the authors and should not be construed as representing the official views or policies of the US Government.
\end{acknowledgments}

\onecolumngrid

\appendix

\section{Quantum state preparation for the H$_2$ molecule}
\label{appa}

To prepare the quantum state in Eq.~\ref{QSP1}, we first construct the compressed state on the first two qubits ($q_1$, $q_2$) with the desired weights  $w_0,w_1,..$. A simple rotation (control-rotation) achieves this. Indeed, note that we need $3$ rotation angles to perform all possible 4-dimensional real vectors $\bd{w}$. If, in addition, we impose the decreasing order $w_j > w_{j+1}$, we can reduce the circuit and rotation angles we used. The preparation of the compressed state is the following:
\begin{align*}
      \ket{0_{q_1}0_{q_2}} &\xrightarrow[]{R_{y} }   \cos{\frac{\theta_0}{2}} \ket{0_{q_1}0_{q_2}} +
    \sin{\frac{\theta_0}{2}} \ket{1_{q_1}0_{q_2}}\\
     &  \xrightarrow[]{C_0R_{y}}
     \cos{\frac{\theta_0}{2}} \cos{\frac{\theta_1}{2}}\ket{0_{q_1}0_{q_2}} + \cos{\frac{\theta_0}{2}} \sin{\frac{\theta_1}{2}}\ket{0_{q_1}1_{q_2}}   + \sin{\frac{\theta_0}{2}} \ket{1_{q_1}0_{q_2}} \\
     &  \xrightarrow[]{C_1R_{y}} \cos{\frac{\theta_0}{2}} \cos{\frac{\theta_1}{2}}\ket{0_{q_1}0_{q_2}} + \cos{\frac{\theta_0}{2}} \sin{\frac{\theta_1}{2}}\ket{0_{q_1}1_{q_2}} + \sin{\frac{\theta_0}{2}}\cos{\frac{\theta_2}{2}} \ket{1_{q_1}0_{q_2}} + \sin{\frac{\theta_0}{2}}\sin{\frac{\theta_2}{2}} \ket{1_{q_1}1_{q_2}} \\
     &= \sqrt{w_0} \ket{0_{q_1}0_{q_2}} +\sqrt{w_1} \ket{0_{q_1}1_{q_2}}+\sqrt{w_2} \ket{1_{q_1}0_{q_2}}+\sqrt{w_3} \ket{1_{q_1}1_{q_2}}\,.
\end{align*}
Next, one can map it back to our desired state \ref{QSP1} using an X and CNOT gates, as indicated in  Figure~\ref{Fig:q_circ}. Assuming all other qubits ($q_2$,$q_3$,$a_0$,$a_1$) are initialized to $0$ we have:
\begin{align*}
     &\sqrt{w_0} \ket{0_{q_1}0_{q_2}} +\sqrt{w_1} \ket{0_{q_1}1_{q_2}}+\sqrt{w_2} \ket{1_{q_1}0_{q_2}}+\sqrt{w_3} \ket{1_{q_1}1_{q_2}} \\
    & \qquad \,\, \xrightarrow[]{X_{q_3}X_{q_4}} \sqrt{w_0} \ket{0_{q_1}0_{q_2}1_{q_3}1_{q_4}} +\sqrt{w_1} \ket{0_{q_1}1_{q_2}1_{q_3}1_{q_4}}  +\sqrt{w_2} \ket{1_{q_1}0_{q_2}1_{q_3}1_{q_4}}+\sqrt{w_3} \ket{1_{q_1}1_{q_2}1_{q_3}1_{q_4}} \\
    &\xrightarrow[{\rm C}_1{\rm NOT}(q_1,a_1)]{{\rm C}_1{\rm NOT}(q_2,a_2)} \sqrt{w_0} \ket{0_{q_1}0_{q_2}1_{q_3}1_{q_4}0_{a_1}0_{a_2}} +\sqrt{w_1} \ket{0_{q_1}1_{q_2}1_{q_3}1_{q_4}0_{a_1}1_{a_2}} +\sqrt{w_2} \ket{1_{q_1}0_{q_2}1_{q_3}1_{q_4}1_{a_1}0_{a_2}}+\sqrt{w_3} \ket{1_{q_1}1_{q_2}1_{q_3}1_{q_4}1_{a_1}1_{a_2}} \\
    &\xrightarrow[{\rm C}_1{\rm NOT}(a_2,q_4)]{{\rm C}_1{\rm NOT}(a_1,q_3)} \sqrt{w_0} \ket{0_{q_1}0_{q_2}1_{q_3}1_{q_4}0_{a_1}0_{a_2}} +\sqrt{w_1} \ket{0_{q_1}1_{q_2}1_{q_3}0_{q_4}0_{a_1}1_{a_2}} +\sqrt{w_2} \ket{1_{q_1}0_{q_2}0_{q_3}1_{q_4}1_{a_1}0_{a_2}}+\sqrt{w_3} \ket{1_{q_1}1_{q_2}0_{q_3}0_{q_4}1_{a_1}1_{a_2}} \\
    &\qquad\,\,\xrightarrow[]{\text{Reorder}} \sqrt{w_0} \ket{1100}_{q}\ket{00}_{a}  +
    \sqrt{w_1} \ket{1001}_{q}\ket{01}_{a} \nonumber + \sqrt{w_2} \ket{0110}_{q}\ket{10}_{a}  +\sqrt{w_3} \ket{0011}_{q}\ket{11}_{a} =\ket{\Phi_{{\rm H}_2}(\bm{w})}\,,
\end{align*}
which is the initial state \eqref{QSP1}.

\section{Calibration of the IBMQ MANILA quantum computer}\label{calibration}
Error channels can be conveniently modeled through Kraus operators, which define a completely positive trace-preserving map on the density matrix:
\begin{equation*}
    \rho \longmapsto \sum_i K_i \rho K_i^\dagger \text{ such that} \sum_i K_i^\dagger K_i =1.
\end{equation*}
In this work, we consider the single-qubit depolarizing error modeled by the following Kraus matrices:
\begin{equation*}
    K_0 =\sqrt{1-p} \, \mathrm{I}, K_{1,2,3}= \sqrt{\frac{p}{3}} \,\sigma_{1,2,3}
\end{equation*}

In this section, we present the parameters of the error model used for running the noisy simulations (see Table \ref{table-ibm}). They were extracted from the ibmq$\_$manila noise model of the Aer Simulator. All the calibration data also can be found in Ref.~\cite{github_code}.

\begin{table*}[t]
\begin{ruledtabular}
\begin{tabular}{lccccc}
Qubit                                    & $Q_0$                                            & $Q_1$                                               & $Q_2$                                               & $Q_3$                                               & $Q_4$                \\ \hline
$T_1$ in $\mu s$                         & 20.931                                          & 145.271                                             & 74.739                                             & 188.731                                             & 138.941               \\
$T_2$ in $\mu s$                         & 18.130                                          & 63.372                                              & 17.264                                              & 60.765                                              & 37.822               \\
Frequency in GHz                         & 4.962                                            & 4.838                                               & 5.037                                               & 4.951                                               & 5.065                \\
Anharmonicity in GHz                     & -0.34463                                         & -0.34528                                            & -0.34255                                            & -0.34358                                            & -0.34211             \\
SPAM error $\ket{1} \rightarrow \ket{0}$ & 0.0574                                           & 0.0556                                              & 0.0338                                              & 0.0384                                               &  0.034               \\
SPAM error $\ket{0}\rightarrow \ket{1}$  & 0.016                                           & 0.0134                                              & 0.0198                                              & 0.0106                                              & 0.0100               \\
Identity,$\sqrt{X}$, X errors ($\times 10^4$)           & 6.2809                                           & 2.1191                                              & 2.5059                                              & 2.0525                                              & 3.8590               \\
CNOT error                               & \multicolumn{1}{l}{0 $\leftrightarrow$ 1: 0.0076} & \multicolumn{1}{l}{1 $\leftrightarrow$ 2 : 0.04386} & \multicolumn{1}{l}{2 $\leftrightarrow$ 3 : 0.03558} & \multicolumn{1}{l}{3 $\leftrightarrow$ 4 : 0.00628} & \multicolumn{1}{l}{} \\
CNOT Gate time in ns                     & 277.333                                          & 469.333                                             & 355.556                                             & 334.222                                             &
\end{tabular}
\end{ruledtabular}
\caption{Calibration used for the ibmq$\_$manila quantum computer.}
\label{table-ibm}
\end{table*}

\section{Optimal sampling in weighted SSVQE}
\label{appd}

For $K$ non-zero weights $w_0 \geq \ldots \geq w_{K-1}$ and $w_0+ \cdots + w_{K-1} =1$, the standard version of SSVQE refers to an initial computational (orthonormal) basis $\{|\phi_0\rangle, ..., |\phi_{K-1}\rangle\}$. A parameterized unitary $U(\bd{\theta})$ is optimized such that its application to these states yields a good approximation to the lowest $K$ eigenstates ($\ket{\phi_j(\bd{\theta})}=U(\bd{\theta}) \ket{\phi_j}$) of a given Hamiltonian $\hat H$. In particular, for a given set of parameters $\bd{\theta}$, the energy expectations $E_j(\bd{\theta}) = \bra{\phi_j(\bd{\theta})}\hat H\ket{\phi_j(\bd{\theta})}$ shall be determined accurately through $M_j$ independent copies of the state $\ket{\phi_j}$ for various $j=0,1,\ldots, K-1$. From a general point of view, there are different ways to choose those various $M_j$. For example, (a) an equal number of measurements $M_0=\ldots = M_{K-1}$, (b) distinctive $M_j$ such that all quantities $E_j(\bd{\theta})$ have the same (relative) error, or (c) optimize
the number of samples to achieve some global energy error.

The sample variance of the individual energy measurements reads
\begin{align}
    \sigma_j^2(\bd{\theta}) = \bra{\phi_j(\bd{\theta})}\hat H^2\ket{\phi_j(\bd{\theta})}- \bra{\phi_j(\bd{\theta})}\hat H\ket{\phi_j(\bd{\theta})}^2
\end{align}
and is related to the estimated energy error of the corresponding state:
\begin{align}
    \varepsilon_j^2 = \frac{\sigma_j^2(\bd{\theta})}{M_j} \,.
\end{align}
The important question is how the total error of the weighted energy, which due to the uncorrelated nature of the statistical fluctuations scales as
$\varepsilon^2 = \sum^{K-1}_{j=0} w_j^2 \varepsilon^2_j$, is related to the total number of samples $M = \sum^{K-1}_{j=0} M_j$ for the three different sampling strategies discussed above.

Choosing an equal number of samples $M_j \equiv M^{(a)}/K$, where $M^{(a)}$ is the total number of samples following strategy (a), will result in the following total error
\begin{align}
    (\varepsilon^{(a)})^2 = \sum^{K-1}_{j=0} w_j^2 \varepsilon^2_j  =   \frac{K}{M^{(a)}} \sum^{K-1}_{j=0} w_j^2 \sigma_j^2(\bd{\theta}) \,.
\end{align}
In contrast, choosing equal individual energy errors $\varepsilon_j = \varepsilon'$, as in strategy (b), results in a total energy error $(\varepsilon^{(b)})^2 = \varepsilon'^2 \sum^{K-1}_{j=0} w_j^2$. Hence,
\begin{align}
    M^{(b)} = \sum^{K-1}_{j=0} M_j = \sum^{K-1}_{j=0} \frac{\sigma_j^2(\bd{\theta})}{\varepsilon^2_j} =  \frac{1}{(\varepsilon^{(b)})^2} \left(\sum^{K-1}_{j=0} w_j^2 \right)\left(\sum^{K-1}_{j=0} \sigma_j^2(\bd{\theta})\right)\,.
\end{align}
The last strategy to execute SSVQE is to choose $M_j$ in order to minimize the total energy error. Following Ref.~\cite{Rubin_2018} we minimize the following Lagrangian:
\begin{align}
    \mathcal{L} = \sum^{K-1}_{j=0} M_j + \lambda \left(\sum^{K-1}_{j=0} \frac{w_j^2 \sigma^2_j(\bd{\theta})}{M_j} -  \varepsilon^2 \right)\,,
    \label{err}
\end{align}
The derivative of $\mathcal{L}$ with respect to $M_j$ gives
$M_j = w_j \sigma_j(\bd{\theta}) \sqrt{\lambda}$.
By replacing this result in the equation for the total error \eqref{err}, one gets the optimal number of total shots:
\begin{align}
M^{(c)} = \frac{1}{(\varepsilon^{(c)})^2}\bigg[\sum^{K-1}_{j=0} w_j \sigma_j(\bd{\theta})\bigg]^2.
\end{align}
We can now compare the three different sampling strategies. For instance, to achieve the same total error (i.e., $\varepsilon^{(a)} = \varepsilon^{(b)} = \varepsilon^{(c)}$), the needed total number of shots of strategies (a), (b) and (c) are related as follows:
\begin{align}
    M^{(c)} =   \frac{M^{(a)}}{K f(\bd{w},\bd{\theta}) }\qquad {\rm and}  \qquad   M^{(c)} =   \frac{M^{(b)}}{g(\bd{w},\bd{\theta}) }\,,
\end{align}
where
\begin{align}
  f  (\bd{w},\bd{\theta}) = \frac{\sum_j w^2_j \sigma^2_j(\bd{\theta})}{\big[\sum_j w_j \sigma_j(\bd{\theta})\big]^2} \qquad {\rm and}  \qquad
  g(\bd{w},\bd{\theta}) = \frac{\big[\sum_j w^2_j\big] \big[\sum_j \sigma^2_j(\bd{\theta})\big]}{\big[\sum_j w_j \sigma_j(\bd{\theta})\big]^2}.
\end{align}
This result indicates that the computational cost depends both on the weights and the quantum circuit parameters.

Now, by using the Cauchy-Schwarz inequality $\big(\sum_j u_j v_j\big)^2 \leq \big(\sum_j u^2_j \big)\big(\sum_j v_j^2\big)$ with $u_j = w_j \sigma_j(\bd{\theta})$ and $v_j =1$ we have
\begin{align}
   \left[\sum^{K-1}_{j=0} w_j \sigma_j(\bd{\theta})\right]^2 \leq  K \sum^{K-1}_{j=0} w^2_j \sigma^2_j(\bd{\theta})\,.
\end{align}
Thus, $K f(\bd{w},\bd{\theta})\geq 1$ and $M^{(c)} \leq M^{(a)}$. A similar reasoning gives $g(\bd{w},\bd{\theta})\geq 1$ and $M^{(c)} \leq M^{(b)}$. To make a further investigation of the terms, we can still assume that all variances are approximately equal. Thus, after using the fact that $\sum_j w_j = 1$, we obtain $f(\bd{w},\bd{\theta}) = \sum_j w^2_j$ and
\begin{align}
    M^{(a)} = M^{(b)} =   \left( K \sum^{K-1}_{j=0} w^2_j\right)  M^{(c)}.
\end{align}
It is easy to show that the prefactor on the right-hand side is bounded from below and above $1 \leq  K \sum^{K-1}_{j=0} w^2_j \leq K$. The lower (upper) bound is attached when all weights are equal $w_j = 1/K$ ($w_0=1$ and $w_{j\neq 0}=0$).

In practice, however, since the individual variances are in general unknown, optimal sampling is achievable either by sampling each state $\ket{\phi_j(\bd{\theta})}$ with probability $w_j$, what we call \textit{importance sampling} SSVQE, or by directly using the cost function Eq.~\eqref{wfield}.  Thus, the total number of samples needed for computing the expected value of the Hamiltonian in the ensemble state up to accuracy $\varepsilon$ is $M = {\rm Var}_{\bm{w}}(H)/\varepsilon^2$, where
\begin{align}
    {\rm Var}_{\bm{w}}(H)  =\sum^{K-1}_{j=0} w_j \bra{\phi_j}(\hat H -\langle \phi_j|\hat H \ket{\phi_j})^2\ket{\phi_j},
\end{align}
which is the same statistical variance of measuring the energy by sampling the pure state~\eqref{wfield}. 

\twocolumngrid

\bibliography{main2}

\end{document}